# Drivers of Layered Circulations in the South China Sea: Volume Flux or Vorticity Flux?

Lei Han [a, b]

[a] *China-ASEAN College of Marine Sciences, Xiamen University Malaysia, Sepang, Malaysia*

[b] *College of Ocean and Earth Sciences, Xiamen University, Xiamen, China*

*Corresponding author*: Lei Han, lei.han@xmu.edu.my

## ABSTRACT

The South China Sea (SCS) features a unique layered circulation structure known as the cyclonic–anticyclonic–cyclonic (CAC) pattern, whose driving mechanisms remain a subject of debate. This study employs idealized numerical simulations using the MITgcm to investigate the role of volume flux and vorticity flux through the Luzon Strait (LS) to the SCS circulation. The control experiment successfully reproduces the observed "sandwiched" transport structure at the LS and the CAC circulation within the SCS. Sensitivity experiments, in which planetary vorticity is set to zero in the Pacific and LS, reveal that the CAC structure is insensitive to vorticity flux input. By applying the Stommel-Arons model to Cartesian coordinates, we provide analytical solutions that align with the simulation results, supporting the volume flux-driven hypothesis. Our findings suggest that the deep inflow through the LS is the primary driver of the layered circulation system, while the middle-layer outflow is a dynamical consequence. These results challenge the vorticity flux-driven hypothesis and offer new insights into the dynamics of layered circulations in the SCS.

## SIGNIFICANCE STATEMENT

The South China Sea (SCS) is a vast body of water linking the Pacific and Indian Oceans. Within the SCS, ocean currents form a distinctive three-layer circulation pattern: flowing counterclockwise near the surface and in the deep layer, but clockwise in the middle. Scientists have proposed two main explanations for this pattern—one attributes it to the volume of water flowing through the Luzon Strait (volume flux), while the other emphasizes the rotational properties of the water (vorticity flux). In this study, we used computer simulations to test both ideas. We found that the deep water entering the SCS through the Luzon Strait—rather than its rotational characteristics—is the primary driver of this layered circulation. This finding enhances our understanding of circulation in the SCS and offers insights that may be relevant to other semi-enclosed seas around the world.

# 1. Introduction

This study investigates the circulation dynamics of the South China Sea (SCS), a tropical deep marginal sea situated between the Pacific and Indian Oceans. The SCS features a complex bathymetry with maximum depths exceeding 4,000 meters in its central basin. Water mass properties and circulation patterns of the SCS are greatly impacted by exchanges with surrounding basins through several straits, among which the Luzon Strait (LS), located in the northeast corner of the basin, is the only deep passage (~2,500 m) (Tian et al 2009; Zhao et al 2016). It connects the SCS to the western Pacific, and the transport through this strait is believed to play a pivotal role in shaping the circulation structure of the SCS (e.g., Gan et al 2016; Zhu et al 2019; Cai et al 2020).

The SCS exhibits a distinctive layered circulation, characterized by a cyclonic–anticyclonic–cyclonic (CAC) pattern in the upper, middle, and deep layers, respectively, with layer interfaces at approximately 500–750 m and 1,500–2,000 m (Cai et al 2020). This CAC structure is supported by multiple lines of evidence. Domain-integrated vorticity from observational datasets and numerical models consistently shows a positive–negative–positive distribution across the upper, middle, and deep layers (e.g., Gan et al 2016; Zhu et al 2017; Zhu et al 2019). Additionally, model-derived flow fields reveal a northward-flowing western boundary current (WBC) in the middle layer and a southward-flowing WBC in the deep layer, which, together with the returning interior flows, forming anticyclonic and cyclonic circulations, respectively (e.g., Yuan 2002; Lan et al 2013; Xu and Oey 2014; Shu et al 2014; Lan et al 2015; Li et al 2021; Zhu et al 2022). Most notably, mooring observations further confirm the deep-layer WBC (Zhou et al 2017; Zhou et al 2020). These findings establish the CAC structure as the mean-state circulation pattern of the SCS (see Table 2 of Zhu et al 2019 for a comprehensive review).

What drives the alternating CAC circulation in the SCS? Cai et al (2020) reviewed several hypotheses that emphasize water exchange through the LS. As the only SCS passage deeper than 500 m (Qu 2000; Li et al 2021), the LS enables transport critical to both middle and deep layers of the CAC structure. Consequently, the LS transport is often regarded as the primary driver of the layered circulation, particularly in the lower two layers.

This pivotal role of the LS arises from its distinct layered transport structure. Across the depth of ~2,500 m, the LS features inflows into the SCS in the upper and deep layers, and an outflow in the middle layer. This “sandwiched” pattern, well-documented by extensive mooring observations (e.g., Tian et al 2006; Yang et al 2010; Zhao et al 2014; Zhou et al 2014; Zhang et al 2015; Zhao et al 2016; Ye et al 2019; Zhou et al 2023), underpins the LS’s influence in shaping the CAC circulation.

As summarized in Cai et al (2020), hypotheses concerning the driving mechanism of CAC circulation can be broadly categorized into two perspectives:

1) **Vorticity-flux-driven hypothesis:**

This hypothesis attributes the layer-dependent circulation patterns to the planetary vorticity flux through the LS. Using a highly idealized single-layer numerical model, Yang and Price (2000) derived an integral constraint for the vorticity budget,

$$\sum_{i=1}^{N} \frac{Q_i f_i}{H_i} = -\lambda \oint_C (\boldsymbol{u}_h \cdot \boldsymbol{l}) ds. \quad (1)$$

Here, $Q_i$, $f_i$, and $H_i$ represent the volume transport out of the basin, Coriolis parameter, and mean layer thickness at the $i$-th opening, respectively. Assuming negligible relative vorticity, the LHS of Eq. (1) denotes the net flux of potential vorticity (PV) ($f/H$) flux into or out of the basin, while the RHS denotes the basin-integrated frictional torque along the lateral boundary, where $\lambda$ is the Rayleigh friction coefficient, $\boldsymbol{u}_h$ the horizontal current velocity, and $\boldsymbol{l}$ the unit vector tangential to the lateral boundary of the basin (positive in counterclockwise direction).

Applied to the SCS, for the deep layer, the LS inflow ($Q_i < 0$) contributes a negative LHS term of Eq. (1), which prompts a counterclockwise (cyclonic) circulation in the deep layer. In contrast, the net outflow in the middle layer results in a positive PV flux (as $Q_i > 0$), thereby favoring a clockwise (anticyclonic) circulation according to Eq. (1).

This hypothesis has been tested in the SCS using a general circulation model in which the LS was artificially closed, resulting in a dramatic alteration of the deep circulation (Lan et al 2013). Zhu et al (2017) further applied this theoretical framework to a multi-layer model to interpret CAC circulation in the SCS.

By analyzing vorticity terms from model output, Gan et al (2016) demonstrated that the vorticity balance in the middle and deep layers is primarily governed by the PV flux through LS and the Joint Effect of Baroclinicity And Relief (JEBAR). The JEBAR term represents a pressure torque generated by the interaction between baroclinically induced pressure gradients and variations in layer thickness, typically associated with bottom topography. It enters the barotropic vorticity equation as a Jacobian term (Mertz and Wright 1992),

$$\Omega^{PGF} = J(\chi, h). \tag{2}$$

where $\chi$ denotes the baroclinic pressure potential, and $h$ is the thickness of the layer in which the vorticity balance is evaluated. Similar vorticity structures have also been identified in subsequent studies, including basin-scale analyses using layered or full circulation models (e.g., Quan and Xue 2018; Wang et al 2018; Cai and Gan 2019; Quan and Xue 2019; Cai et al 2023), as well as regional analyses based on in-situ current measurements from mooring arrays in the SCS (Zheng et al 2024).

In summary, this hypothesis posits that the CAC circulation is "largely induced by planetary potential vorticity flux through lateral boundaries, mainly the Luzon Strait" (Zhu et al 2017). The underlying mechanism is that net PV outflow in the middle layer and PV inflow in the upper and bottom layers result in negative and positive layer-averaged vorticity, respectively, giving rise to basin-wide anticyclonic circulation in the middle layer and cyclonic circulation in the upper and deep layers (e.g., Zhu et al 2017; Zhu et al 2019; Cai et al 2020).

**2) Volume-flux-driven hypothesis:**

The volume-flux-driven hypothesis, like the vorticity-flux-driven hypothesis, attributes the CAC circulation to exchange through the LS. However, it differs by emphasizing the role of the volume flux itself, rather than vorticity flux, as the primary driving mechanism.

This hypothesis traces its origins to a theory proposed by Stommel and Arons (1959) over six decades ago. Similar to the single-layer model of Yang and Price (2000), the original Stommel-Arons theory was not intended to explain multilayer circulations; instead, it focused on dynamics of the abyssal circulation and the deep western boundary current (Stommel et al 1958).

The application of the Stommel-Arons theory to the CAC circulation relies on the $\beta$-effect in the SCS. As deep inflow through the LS enters the SCS, it compresses water columns laterally, leading to vertical stretching by continuity. To conserve PV, the stretched water columns tend to move poleward. According to classical gyre dynamics (e.g., Stommel 1948; Munk 1950), an equatorward WBC develops to balance the poleward interior flow, thereby forming a cyclonic circulation in the deep layer.

In contrast, the middle layer experiences vertical compression from below. This vertical squeezing causes the water columns in the middle layer to move equatorward due to PV conservation, resulting in a poleward WBC and an anticyclonic circulation.

The volume-flux-driven hypothesis can be described by the PV balance below (Yuan 2002),

$$f\frac{dw}{dz} = \beta v. \tag{3}$$

where $\beta$ is the meridional gradient of the Coriolis parameter, $f$, and $v$ is the meridional velocity of a water parcel. A key distinction from the vorticity-flux-driven hypothesis is that this hypothesis does not require PV fluxes at lateral boundaries. Using a general circulation model, Yuan (2002) successfully reproduced CAC-like circulations in the SCS, and highlighted the $\beta$-effect as being of “paramount importance” to the system.

Despite its conceptual appeal, the volume-flux-driven hypothesis has received relatively limited attention—likely due to the prevailing view that it "cannot provide the full dynamics of the SCS basin circulation in response to external forcing of the mass source/sink" (Cai et al 2020).

To summarize, which hypothesis better reflects the actual dynamics underlying the CAC circulation remains an open question. Disentangling the two mechanisms in process-oriented studies is inherently challenging, as volume flux and vorticity flux often occur simultaneously. For instance, Yuan (2002) removed the $\beta$-effect in the SCS while retaining it in the Pacific, resulting in a substantial reduction of transport through the LS. However, this reduction affected both volume and vorticity fluxes, making it difficult to isolate their respective impacts. A similar limitation applies to the experiment conducted by Lan et al (2013), which reported a "prominent difference" in circulation structure when the LS was artificially closed.

This study aims to address this question by designing targeted numerical experiments that isolate the two types of forcing at the LS. Unlike the middle and deep layers, the upper-layer circulation is influenced not only by the LS transport but also by surface forcing—particularly wind stress—which leads to pronounced seasonal variability (e.g., Chu et al 1999; Qu 2000; Gan et al 2016; Cai and Gan 2020). To isolate the key dynamical processes governing the layered structure, we deliberately exclude surface forcing from our model setup; consequently, the upper-layer circulation will not be examined in this study.

The remainder of the paper is organized as follows. Section 2 introduces the idealized model configurations used in this study. Section 3 presents the features of the layered transports and circulations in the control run and sensitivity experiments. Section 4 applies the Stommel-Arons theory to the SCS and evaluates the theoretical solutions against the model results. Section 5 provides a summary and discussion.

## 2. Model description

To reproduce the layered circulations in SCS and to assess their sensitivity to key environmental parameters, we employ a process-oriented numerical framework based on the Massachusetts Institute of Technology general circulation model (MITgcm). A box model configuration is used, which has proven valuable in process-oriented studies of global ocean dynamics (e.g., Ito and Marshall 2008; Wolfe and Cessi 2010; Nikurashin and Vallis 2011; Wolfe and Cessi 2011; Munday et al 2013; Bell 2015; Mashayek et al 2015; Ferrari et al 2016). Box models have also been applied to the SCS in studying the CAC dynamics (e.g., Cai and Gan 2019; Cai et al 2023).

The model setup is described below, with the values of key model parameters summarized in Table S1.

**2.1 Domain and topography**

The computational domain consists of two adjacent basins: one representing SCS and the other representing the connected region of the western Pacific Ocean (Fig. 1a). The computation domain of Pacific basin extends 30° in longitude and from the Equator to 35°N in latitude (Fig. 1a, b), similar to the geometry employed by Cai et al (2023) (hereafter CCG23), which covers 25° in longitude and approximately 30° in latitude. The idealized SCS spans approximately 1,000 km in zonal direction and 1,200 km in meridional direction, corresponding to roughly 10°×12°, starting from 10°N (Fig. 1b, c). In this study, a full-width Pacific configuration with a 140° zonal extent (purple box in Fig. 1a) is also tested to assess the sensitivity of the results to the width of the outer basin (see results in Section 3.1).

Given that the SCS is largely enclosed below 600 m except for one major outlet (LS), and that our focus is on the middle and deep layers, we represent the LS as the only channel (350 km wide meridionally) connecting the two model basins. Both basins are assigned a uniform depth of 4000 m, with the sill depth at LS set to 2000 m (Fig. 1b, c). To our knowledge, no prior studies have applied flat-bottom topography in the SCS to investigate layered gyre circulations, highlighting the importance of exploring the dynamical response of SCS circulation in the absence of topographic effects.

The model employs a horizontal resolution of 50 km and a vertical resolution of 80 m. To address concerns that 50 km may be insufficient to resolve circulation features in the SCS, we also conduct a 10 km (i.e., 0.1°) resolution experiment to test the robustness of the results at eddy-resolving scales (see results in Section 3.1).

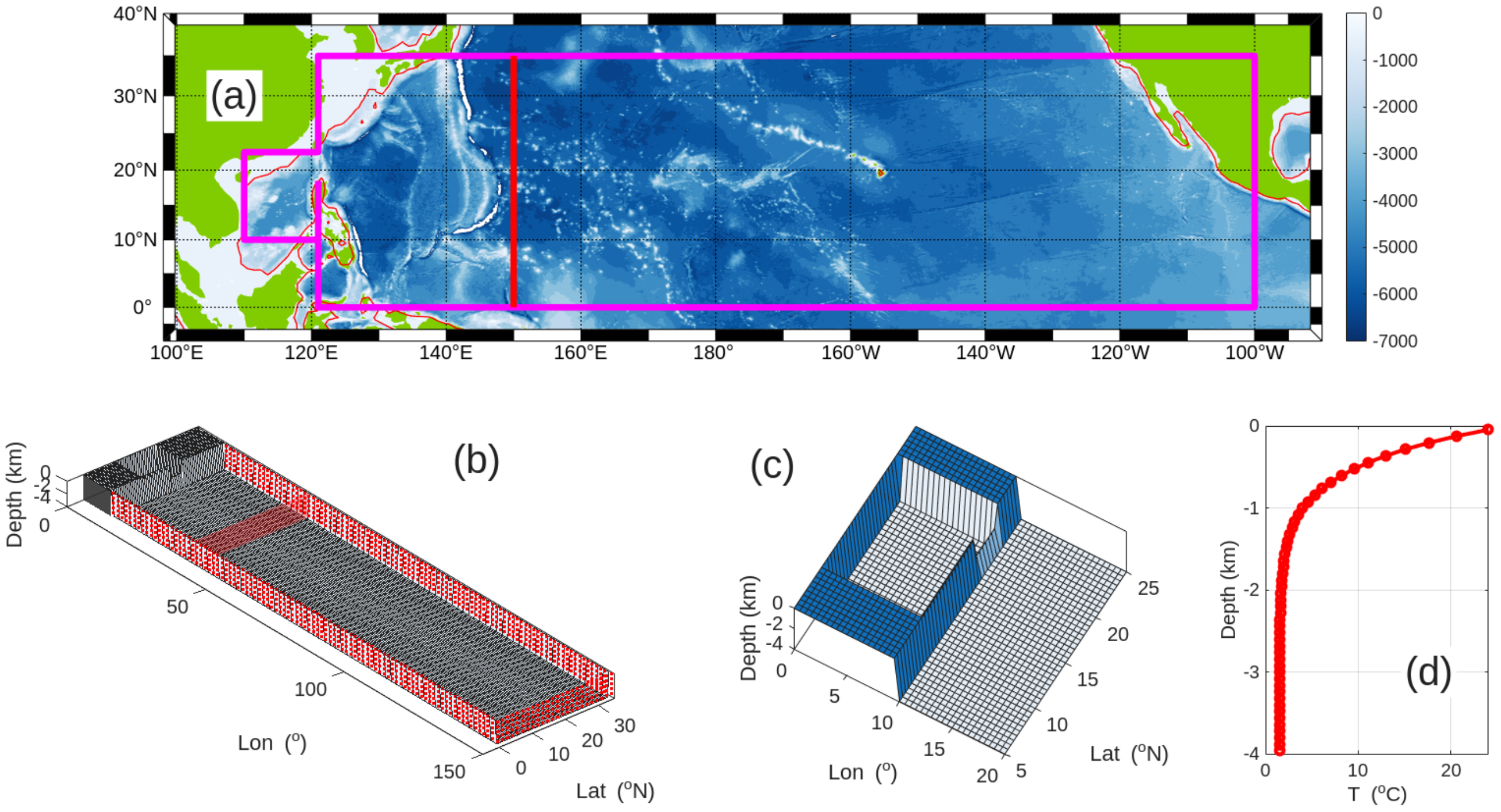

Fig. 1. (a) Bottom topography of the South China Sea (SCS) and its adjacent region in the Pacific Ocean. The red curve marks the 600 m isobath. The area enclosed by the purple indicates the computational domain for the full-width Pacific configuration, whose 3D view is shown in (b). The control and sensitivity experiments adopt a reduced domain, terminating at 150°E in (a) (red thick line) or at 40° longitude in (b) (red meridional section). Temperature is relaxed toward the initial temperature field at the outer boundary of the Pacific basin, as indicated by the red dotted and red surface areas in (b). A zoomed-in view of the western portion of the domain is shown in (c), and the initial temperature profile is presented in (d).

## 2.2 Initial conditions

All simulations, including the control run and the sensitivity experiments, are initiated from a state of rest. Following common practice in similar studies (e.g., Ferrari et al 2016), salinity is held constant throughout, so that potential density depends solely on temperature. A linear equation of state is employed. The initial temperature field is horizontally uniform and vertically stratified, decreasing exponentially from 24 °C at the surface to 1.5 °C at the bottom (Fig. 1d).

The Coriolis parameter is zero at the Equator, and varies linearly with latitude under the $\beta$-plane approximation for the control run. A variety of Coriolis parameter distribution is tested in the sensitivity experiments, as introduced in Section 2.4.

### 2.3 Forcing conditions

To isolate the impacts of inter-basin exchange through the LS and minimize the influence of other factors, we adopt a minimal-forcing strategy:

1) **Surface forcing: None**

   As this study focuses on the mid- and deep-layer circulations, all surface forcings (e.g., wind stress, heat flux) are turned off. This modeling choice is supported by previous numerical studies that successfully reproduced the CAC structure in the absence of wind forcing (e.g., Yuan 2002; Cai and Gan 2019; CCG23).

2) **Lateral temperature relaxation:**

   To achieve steady-state solutions, temperatures along the three boundaries of the Pacific basin (the red dotted/surface in Fig. 1b) are relaxed monthly toward the initial field. This boundary treatment maintains a stable external environment for the SCS, consistent with the idealized approach of CCG23 (personal communication with Z. Cai, 2025).

3) **Diapycnal diffusivities ($\kappa$):**

   The key driver of exchange across the LS is the contrast in mixing intensity between the two basins. Observations indicate that diapycnal diffusivity in the SCS is approximately two orders of magnitude greater than in the western North Pacific (e.g., Tian et al 2009; Alford et al 2011; Yang et al 2016), due to enhanced internal-wave activity (Tian et al 2009; Yang et al 2016; Wang et al 2017). This mixing contrast sustains the density gradient between the SCS and Pacific, which in turn drives cross-strait transport (e.g., Wang et al 2011; Zhao et al 2014; Zhou et al 2014; Wang et al 2017). Previous studies have shown that increasing mixing in the SCS strengthens deep inflow through the LS (e.g., Zhao et al 2014; Wang et al 2017; CCG23).

   Following these findings, we prescribe two uniform diapycnal diffusivities:

$\kappa$=1×10$^{-3}$ m$^2$/s in the SCS and LS,

and $\kappa$=1×10$^{-5}$ m$^2$/s in the Pacific basin.

While this uniform setup differs from earlier studies that employed depth-dependent diffusivity profiles (e.g., Zhao et al 2014; Wang et al 2017; CCG23), we deliberately adopt the simplified form to isolate and clarify the underlying dynamics. We acknowledge that vertical variations in $\kappa$ can affect the vertical exchange and, in turn, the horizontal circulations. Nevertheless, the uniform-$\kappa$ configuration offers a useful baseline for understanding the fundamental processes and serves as a foundation for future studies involving more realistic, vertically varying mixing schemes.

**2.4 Sensitivity test configuration**

As discussed in Section 1, testing the two hypotheses regarding the formation of the CAC circulation in the SCS requires idealized experiments that can isolate the volume flux and vorticity flux through the LS. Yuan (2002) demonstrated that the water exchange through the LS is highly sensitive to the Coriolis parameter $f$ within the SCS. However, that study only considered a constant-$f$ case in the SCS and did not explore scenarios with $f$=0 in the Pacific and LS.

To address this gap, we design three sensitivity experiments to evaluate how the SCS circulation responds under conditions of zero planetary vorticity input into the basin. In contrast to the control run (hereafter Case-1), which employs a realistic $f$-field throughout the entire domain, all three sensitivity experiments set $f$=0 in both the Pacific domain and the LS, thereby eliminating the influx of planetary vorticity. Within the SCS, different Coriolis configurations are prescribed, as summarized below and illustrated in Fig. 2:

- Case-2: Identical to the control run (Case-1), i.e., $f > 0, \beta > 0$.
- Case-3: $f > 0, \beta > 0$, but with $\beta$ oriented in the zonal direction such that $f$ increases westward;
- Case-4: $f < 0, \beta > 0$.

All four experiments (Cases 1–4) adopt the same magnitude of $\beta$, as specified in Table

S1.

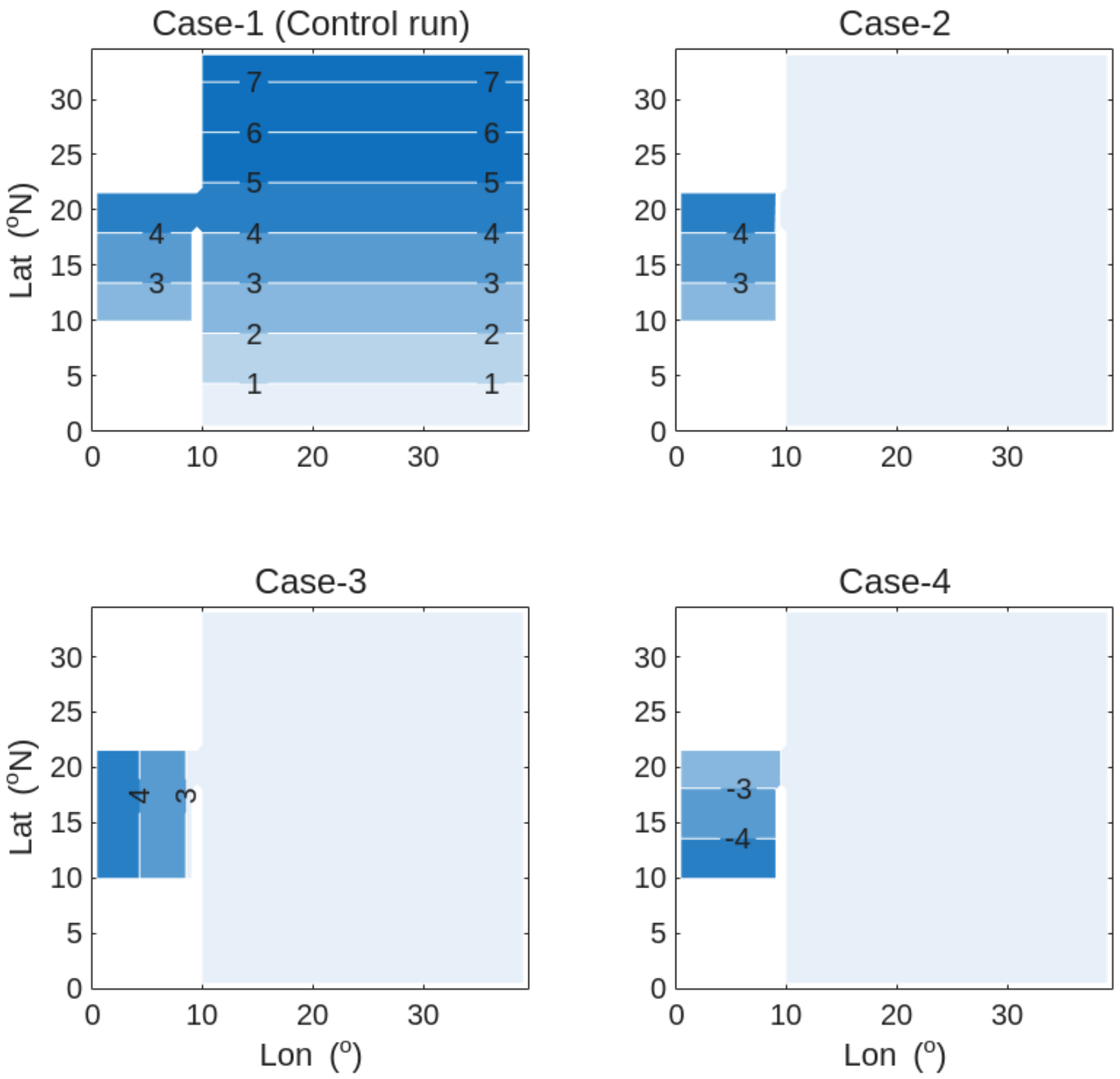

Fig. 2. Distribution of the Coriolis parameter *f* in the control run and the three sensitivity experiments. Contour labels are in units of $1\times10^{-5}$ $s^{-1}$. In Cases 2–4, *f* is set to zero in both the Pacific domain and the LS to preclude the planetary vorticity flux into the SCS.

**2.5 Summary of model configuration**

In addition to precluding planetary vorticity influx, this study introduces several key modeling simplifications that distinguish it from previous work. These simplifications are applied consistently across all experiments and are intended to clarify the fundamental dynamics underlying the CAC circulation. In particular, compared to the recent idealized simulation by CCG23, which successfully reproduced the layered structure in both the LS transport and the SCS circulations, our model incorporates two major simplifications:

**1) Absence of topography in the SCS**

While CCG23 implemented a circular basin with sloping bottom topography, our model assumes a flat-bottom bathymetry in the SCS. In this setup, the absence of a topographic gradient precludes the JEBAR term from Eulerian momentum equation altogether, even when applied in a layered formulation. (e.g., uniform layer thickness below 1500 m). This configuration enables us to assess whether the JEBAR mechanism is indispensable for the emergence of the layered circulation.

2) **Absence of Kuroshio forcing**

In CCG23, inflow/outflow boundary conditions were applied at the eastern boundary, and surface wind stress was imposed over the Pacific domain to generate a realistic WBC. In contrast, our model excludes all external dynamical forcing in both the Pacific and SCS basins, allowing us to isolate and examine whether the inter-basin density contrast alone is sufficient to drive significant layered transport across the LS.

Additional simplifications relative to CCG23 include the use of uniform diffusivity profiles and a coarser horizontal resolution (50 km vs. 5 km). Collectively, these deliberate simplifications form a minimalistic yet dynamically informative framework for identifying the essential physical mechanisms responsible for the CAC circulation.

All simulations were integrated for 200 years to reach a quasi-equilibrium state, and the final 10-year averages are used in the analysis presented in the following sections. Intermediate results from model years 91–100 are also examined to assess how closely the system has approached equilibrium (see results in Section 3.1).

# 3. Model results

## 3.1 Assessment of the control run

We begin by assessing the results of the control run (Case-1), with particular focus on examining the layered features in the LS and SCS basin. These results serve as a benchmark for evaluating the model's performance and for comparison with the sensitivity experiments presented in the next section.

1) **Sandwiched transport structure at the LS**

The sandwiched vertical structure of "inflow–outflow–inflow" through the LS is well reproduced in the control run (Fig. 3). Notably, the middle-layer outflow is concentrated near the southern end of the strait, in good agreement with field observations (Zhang et al 2015; Yang et al 2016). The model-predicted layer interfaces are located at approximately 200 m and 1,200 m, respectively, which compares reasonably well with observation-based estimates of ~500 m and ~1,500 m (e.g., Tian et al 2006; Yang et al 2010), considering that the sill depth of the LS is set to 2,000 m in the model versus ~2,500 m in reality.

The simulated net transports—3.0 Sv for the bottom inflow and 5.6 Sv for the middle outflow—are in good agreement with observational estimates, which range from 2.0–3.0 Sv for the bottom inflow and 2.5–5.0 Sv for the middle outflow (Yang et al 2010). Notably, the simulated transport ratio of the two layers (3.0 / 5.6 = 0.54) falls comfortably within the observed range of 0.4–0.8, lending further confidence in the model's ability to capture the vertical partitioning of the exchange.

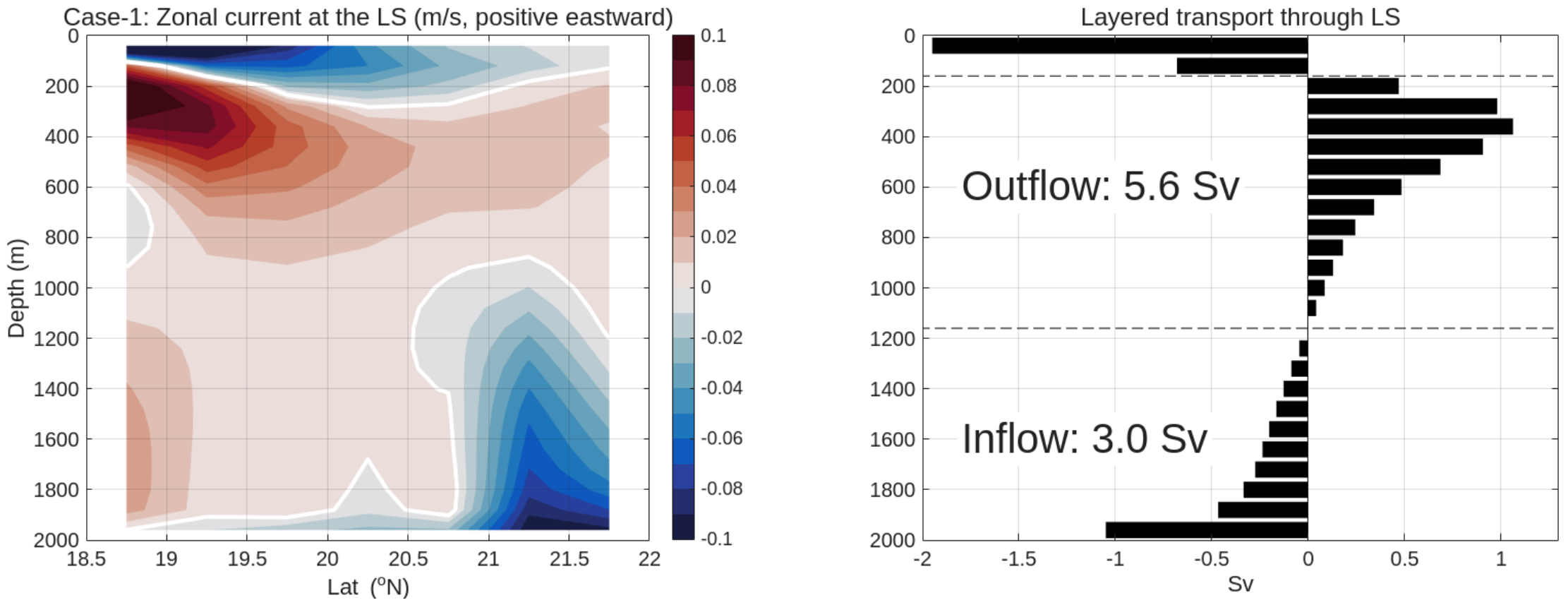

Fig. 3. Zonal currents (left) and layer-integrated zonal transport (right) through the LS in Case-1 (control run). The thick white line in the left panel marks the zero-velocity contour. The characteristic "sandwiched" structure is clearly reproduced. The two dashed lines indicate depths of 160 m and 1,160 m. See text for discussions.

**2) CAC circulations in the SCS basin**

In addition to reproducing the layered transport structure in the LS, the model successfully captures the characteristic CAC circulation pattern within the SCS basin (Fig. 4). Prominent WBCs exhibit a distinct vertical alternation, flowing northward in the middle layer (Fig. 4a, c) and southward in the deep layer (Fig. 4a, d). Together with the opposing return flows in the basin interior, these WBCs give rise to an anticyclonic circulation in the middle layer and cyclonic circulations in the deep layer.

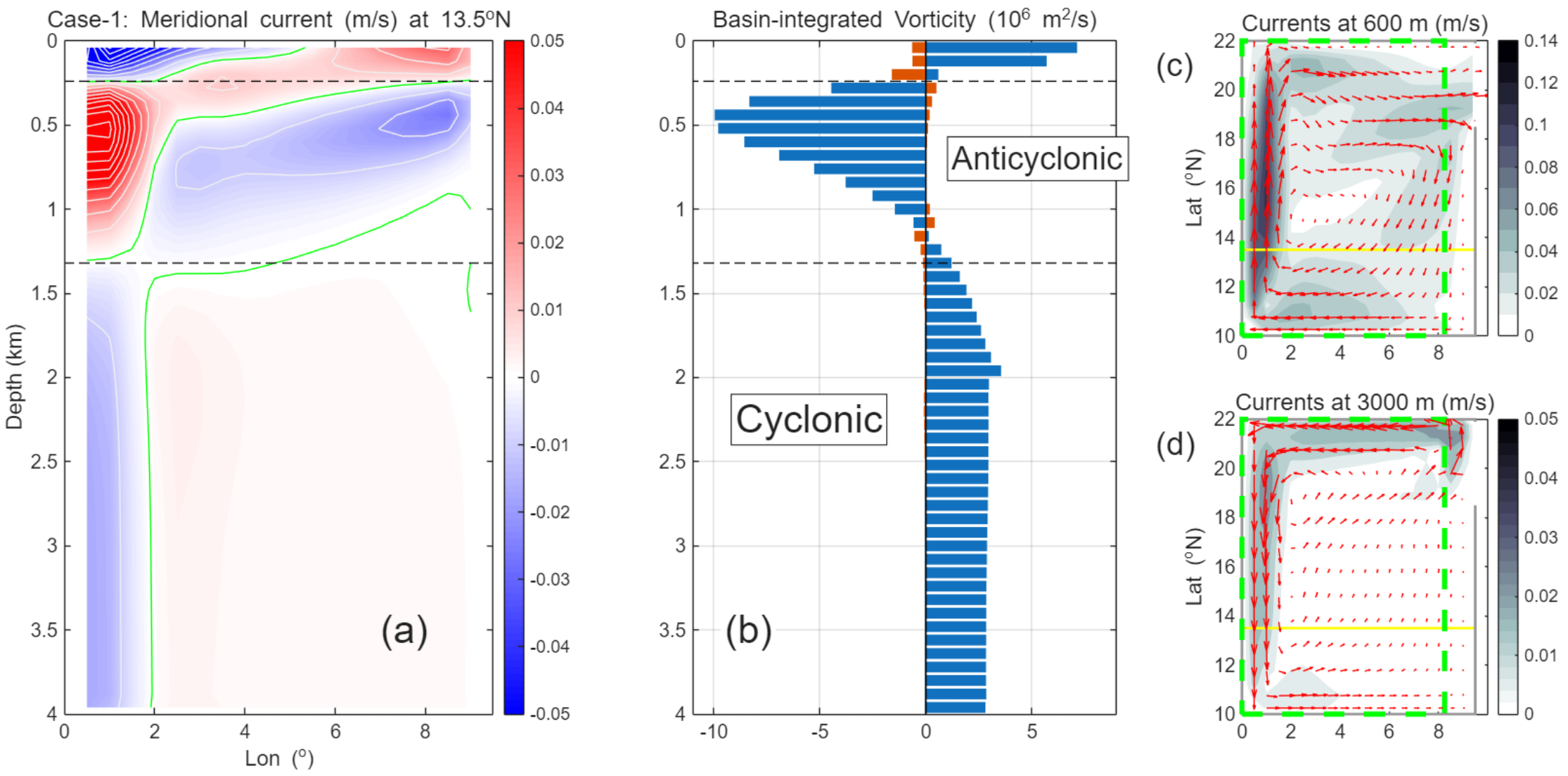

Fig. 4. Modeled circulations in Case 1 within the SCS. (a) Zonal section of the meridional velocity (in m/s) at 13.5°N. Positive (negative) values indicate northward (southward) flow. The zero contour is highlighted in green. Two horizontal dashed lines (240 m and 1,320 m) mark the layer interfaces separating alternating spinning circulations. (b) Basin-integrated vorticity (in $m^2/s$) as a function of depth. Positive (negative) values correspond to cyclonic (anticyclonic) circulation. See text for computational details. Dashed lines indicate the same interface depths as in (a). (c, d) Horizontal velocity vectors (red arrows) and current speed (color shading, in m/s) at 600 m and 3,000 m, respectively. The yellow line indicates the transect location shown in (a), and the green dashed box marks the area over which the basin-integrated vorticity is calculated.

To quantify the direction and strength of these circulations, we calculate the basin-integrated, following the approach adopted in previous studies (e.g., Gan et al 2016; Zhu et al 2017; Cai and Gan 2019; Quan and Xue 2019; Zhu et al 2019; CCG23). According to Stokes' theorem, the domain-integrated vorticity is equivalent to the line integral of tangential velocity along the boundary of the domain. Notably, this relationship holds naturally in the discretized model using the Arakawa C-grid (as in MITgcm) without requiring explicit enforcement.

The depth profile of basin-integrated vorticity is shown as blue bars in Fig. 4b. To further assess the robustness of the vorticity sign (positive or negative) at each depth level, we also compute the contributions from the along-boundary velocities opposing the direction of the total line integral (positive counterclockwise), shown as red bars in Fig. 4b. These opposing contributions indicate uncertainty in the vorticity sign. For example, the vorticity at the 3$^{rd}$ level (~200 m) appears less reliable as the opposing contribution (red bar) exceeds the net value (blue bar), suggesting comparable positive and negative components in the line integral.

Despite local uncertainties, the overall layered pattern of alternating vorticity remains clear and robust, closely aligning with the vertically alternating WBC structure defined by layer interfaces in Fig. 4a and 4b. Notably, basin-integrated vorticity is computed over a subregion of the SCS (green dashed box in Fig. 4c, d), rather than the full domain. This choice ensures consistency with the sensitivity experiments presented in Section 3.2, where strong recirculation near the LS tends to dominate the basin-integrated vorticity and obscure the interior circulation signal (Section 3.2). For the control run, however, similar results are obtained whether the full domain or subregion is used, as the velocities along the eastern boundary are relatively weak (e.g., Fig. 4c, d).

Although the model successfully reproduces the LS inflow and the associated cyclonic circulation in the upper layer (Figs. 3 and 4), consistent with the upper-layer component of the CAC system, this layer is not examined in the present study. This is because surface forcing—likely a key driver of upper-layer circulation—is deliberately excluded to simplify the model dynamics. The upper-layer inflow into the SCS is likely driven by the sea surface height (SSH) difference between the two basins, which closely correlates with the sea surface temperature (SST) distribution (Fig. S1). The relatively colder SST in the SCS is likely a consequence of stronger vertical mixing compared to the Pacific side.

**3) Vertical overturning circulation**

Examining the meridional overturning circulation (MOC) within the SCS basin provides valuable insights into vertical water transport and inter-layer exchange—key processes that shape the layered circulation structure (e.g., Wang et al 2018; Quan and Xue 2019). Here, we compute two types of MOC streamfunctions using utility tools provided in MITgcm: the Eulerian MOC streamfunction (denoted as $\psi_{Eul}$) and the residual MOC streamfunction (denoted as $\psi_{Res}$). The residual MOC reflects the net transport resulting from the combined effects of the Eulerian-mean and eddy-induced circulations (e.g., Marshall and Radko 2003; Nurser and Lee 2004; Ito and Marshall 2008; Munday et al 2013), and closely resembles the Eulerian MOC in much of the interior basin where isopycnals are relatively flat (Ito and Marshall 2008).

In Case-1, both streamfunctions exhibit a similar two-cell structure (Fig. 5). The lower upwelling cell, with a mean strength of ~2 Sv, represents the return pathway of dense water supplied by the deep inflow through the LS, which rises from the deep layer into the middle layer before exiting the basin. The upper downwelling cell, with a strength of ~1.5 Sv, indicates the vertical transformation and subduction of upper-layer water into the middle layer, which subsequently exits through the LS. Together, these overturning patterns reflect the vertical mass redistribution associated with the LS-driven inter-layer exchange. The MOC streamfunctions in Case 1 are consistent with the schematic two-cell overturning structure proposed by Wang et al (2016), lending further support to that conceptual framework.

It is worth noting that the difference between the two MOC streamfunctions near the seafloor is substantially reduced in the higher-resolution (10 km) experiment (see Section 3.2 or Fig. S6).

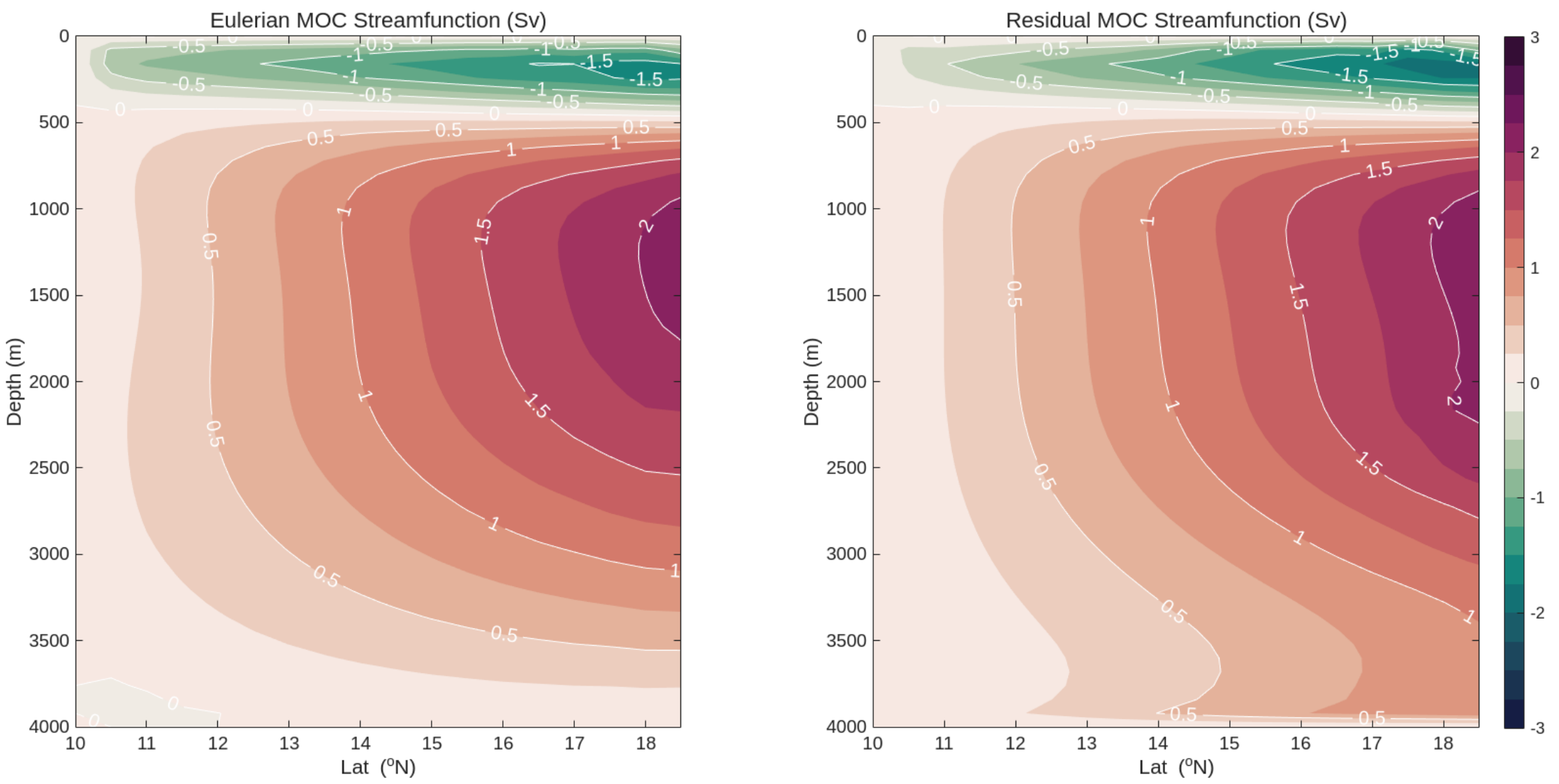

Fig. 5. The Eulerian (left) and residual (right) meridional overturning circulation (MOC) streamfunctions in the SCS basin in Case-1. Positive (negative) values indicate clockwise (counterclockwise) or upwelling (downwelling) overturning cells.

### 4) Testing the robustness of model output

To verify the reliability of the results obtained from the control run (Case-1), we performed additional analyses and experiments to evaluate the model's sensitivity to various settings. Based on the Case-1 configuration, four test cases were designed to examine different aspects as follows:

a) **Case-1/A (Equilibrium status):** Mean values were computed over model years 91–100, rather than 191–200 as in Case-1, to assess how closely the system has reached a steady state.

b) **Case-1/B (Time step sensitivity):** The time step was doubled to 60 minutes (from 30 minutes in Case-1) to examine sensitivity to temporal resolution.

c) **Case-1/C (Pacific width):** Identical to Case-1/B, but with a full-width Pacific domain (120°E–100°W; see Fig. 1) to test the sensitivity to the zonal extent of the outer basin.

d) **Case-1/D (Horizontal resolution):** Identical to Case-1/B, but with a finer horizontal resolution of 10 km (instead of 50 km for Case-1) to evaluate the impact of spatial resolution.

The metrics from these four cases are compared with those of the control run in Table S2 (see Section 3.2 for definitions of the metrics). Across various metrics, the differences are minimal, with the exception of Case-1/D, where the higher resolution produces a notably stronger LS outflow, an enhanced WBC transport, and increased basin-integrated vorticity in the middle layer. Despite these quantitative differences, the qualitative circulation features remain consistent with those in the control run. Circulation patterns for Case-1/C and Case-1/D are illustrated in Figs. S2–S3 and Figs. S4–S6, respectively.

These findings demonstrate that the simulated sandwiched LS transport and the CAC circulations in the SCS are robust in Case-1, showing little dependence on the outer basin width or model resolution.

To conclude Section 3.1, the model performs remarkably well in capturing the essential features of the layered exchange through the LS and the basin-scale circulation patterns in the SCS—despite its coarse resolution, simplified geometry and topography, and minimal external forcing. This strong performance provides confidence in the model's utility as a process-oriented tool for exploring the dynamical drivers of the circulation. In the next section, we analyze a suite of sensitivity experiments, with configurations outlined in Section 2.4.

### 3.2 Three cases with zero influx of planetary vorticity

Building on the successful control run, we investigate the role of planetary vorticity influx through the LS by prescribing different Coriolis parameter distributions within the SCS domain (Cases 2–4; Fig. 2). In all three experiments, the Coriolis parameter is set to zero in both the LS and Pacific regions, effectively eliminating any planetary vorticity input from the Pacific side.

Despite this constraint, all sensitivity experiments reproduce a robust layered structure in both the LS transport and the horizontal circulation within the SCS, with transport magnitudes comparable to those in the control run (Figs. 6–11).

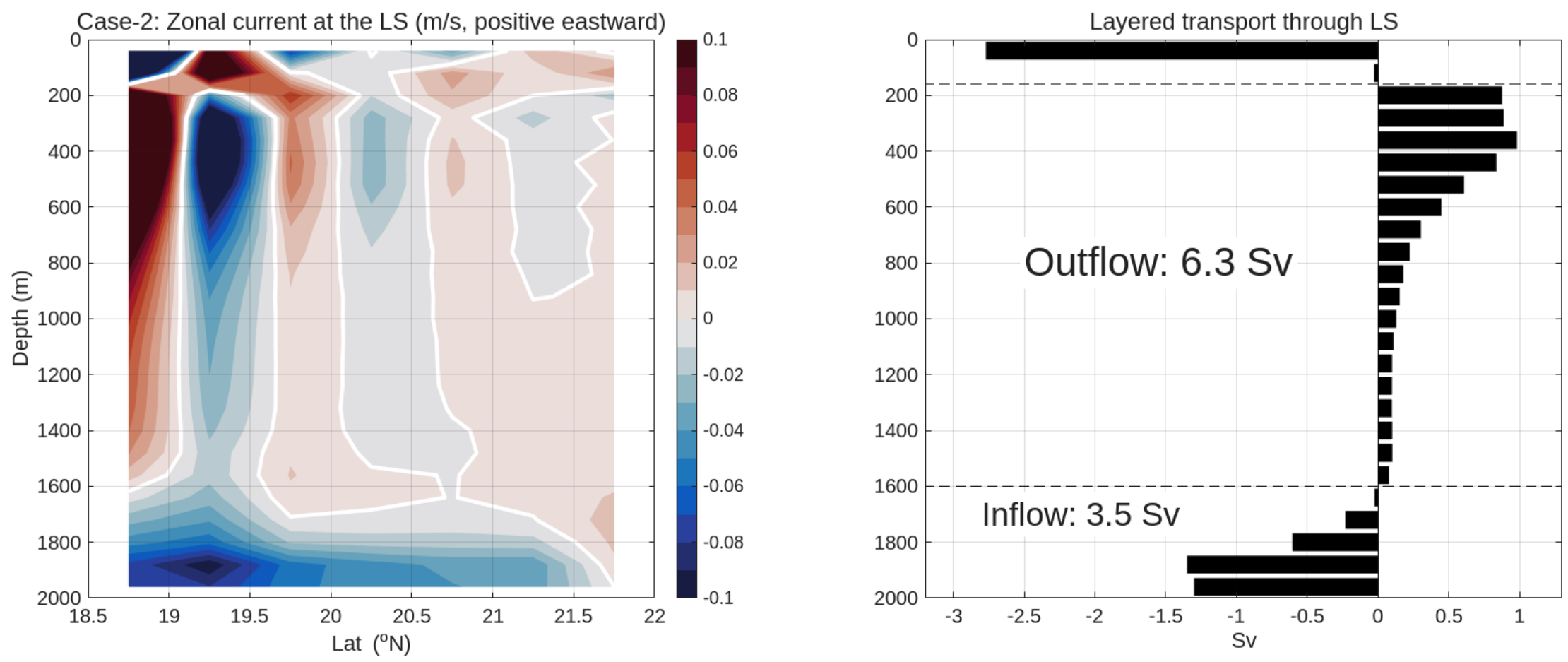

Fig. 6. Same as Fig. 3, but for Case-2.

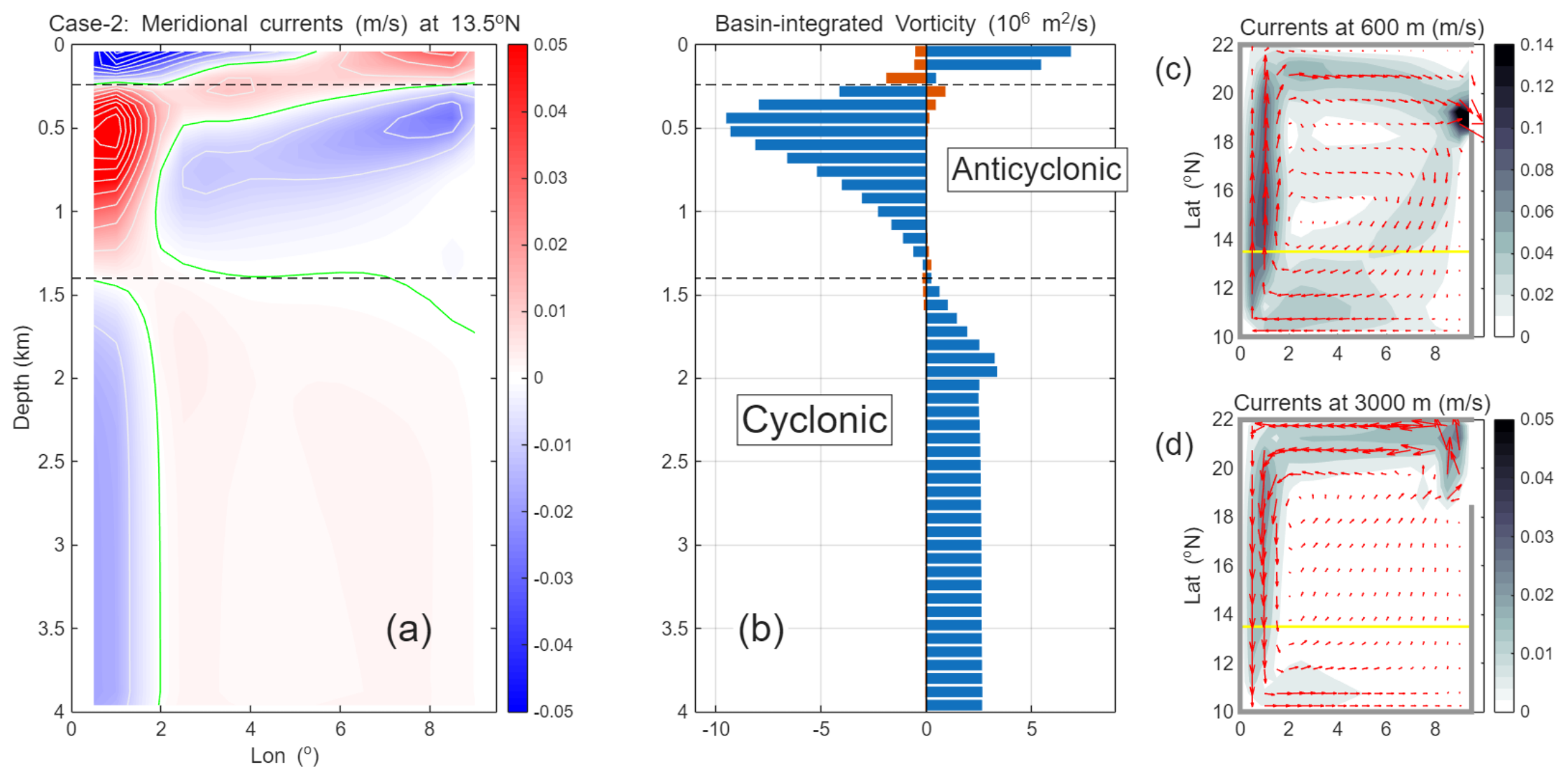

Fig. 7. Same as Fig. 4, but for Case-2.

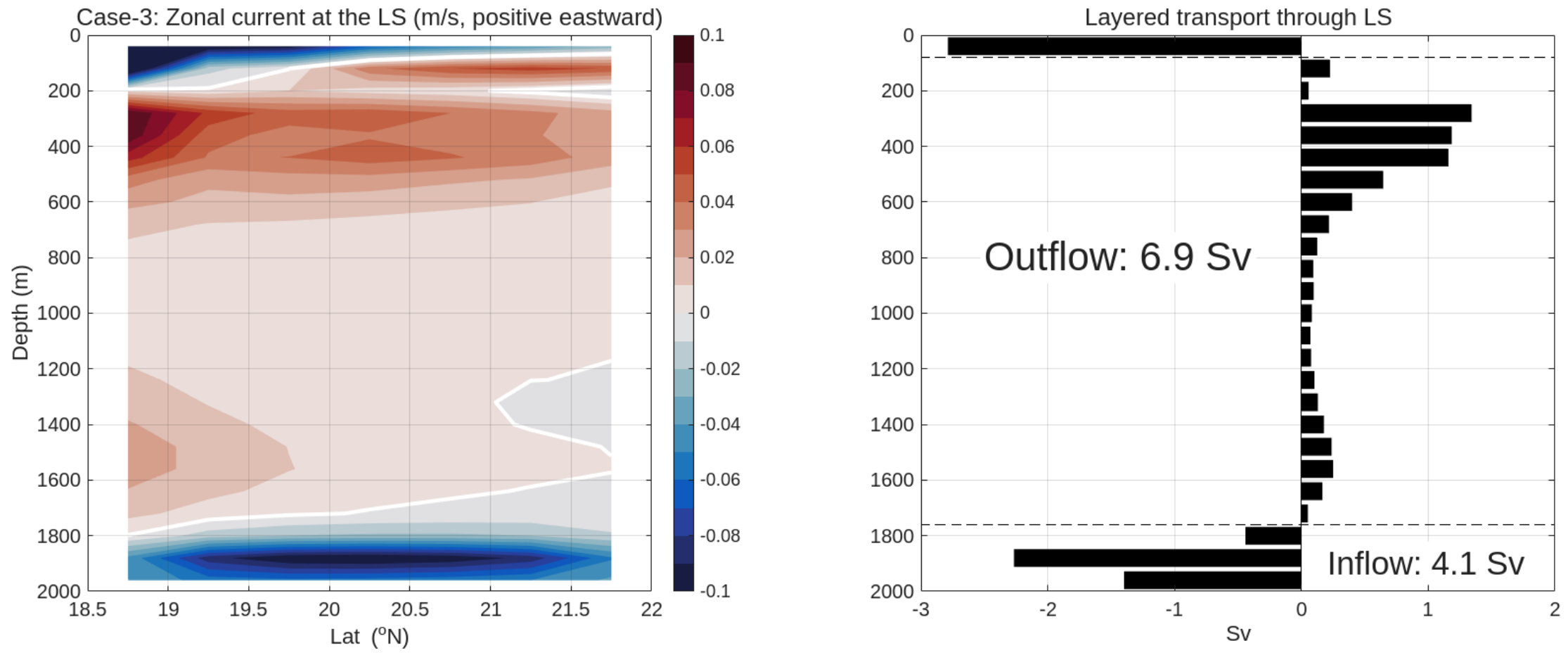

Fig. 8. Same as Fig. 3, but for Case-3.

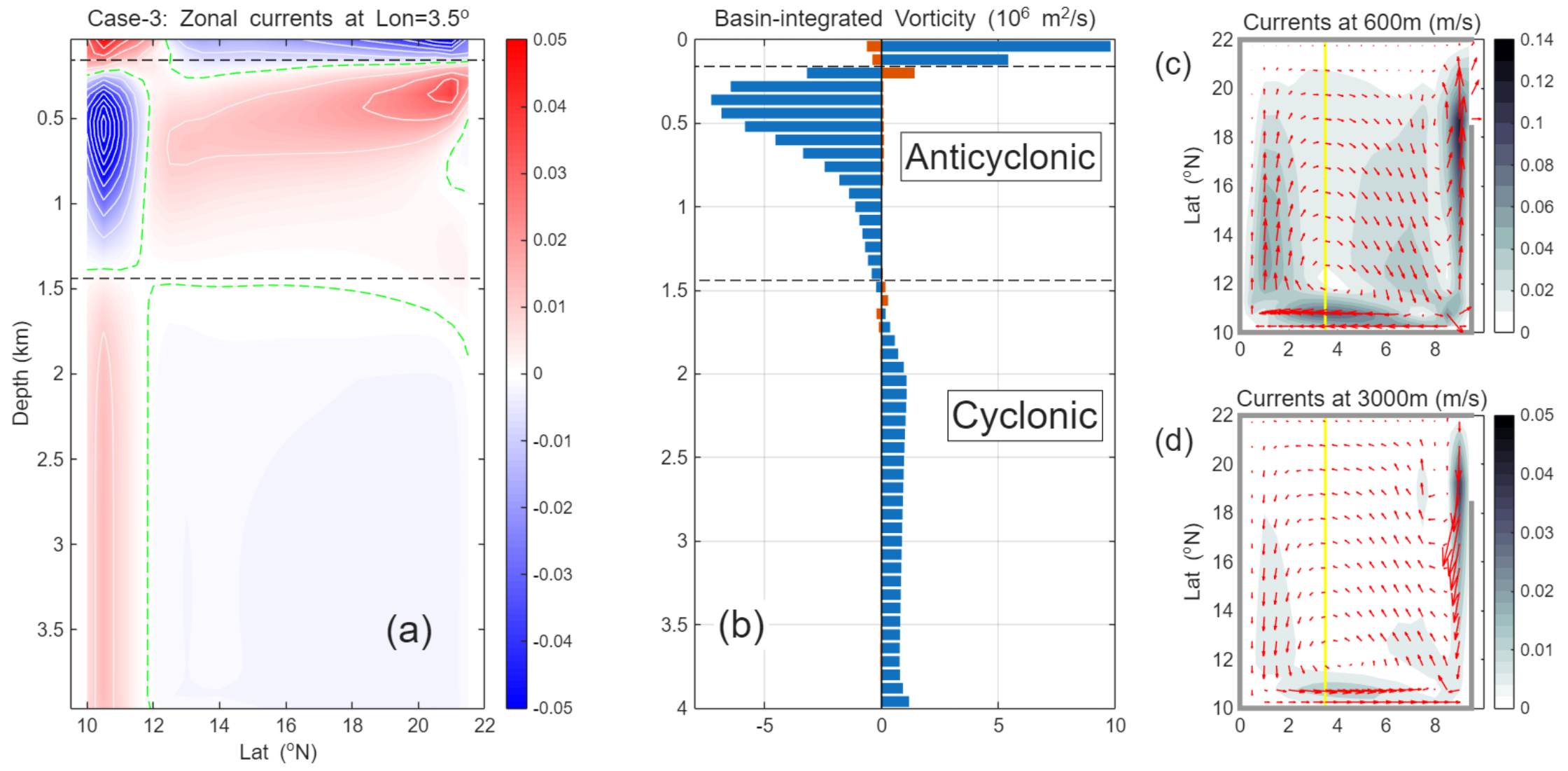

Fig. 9. Same as Fig. 4, but for Case-3, and the section is meridional instead of zonal.

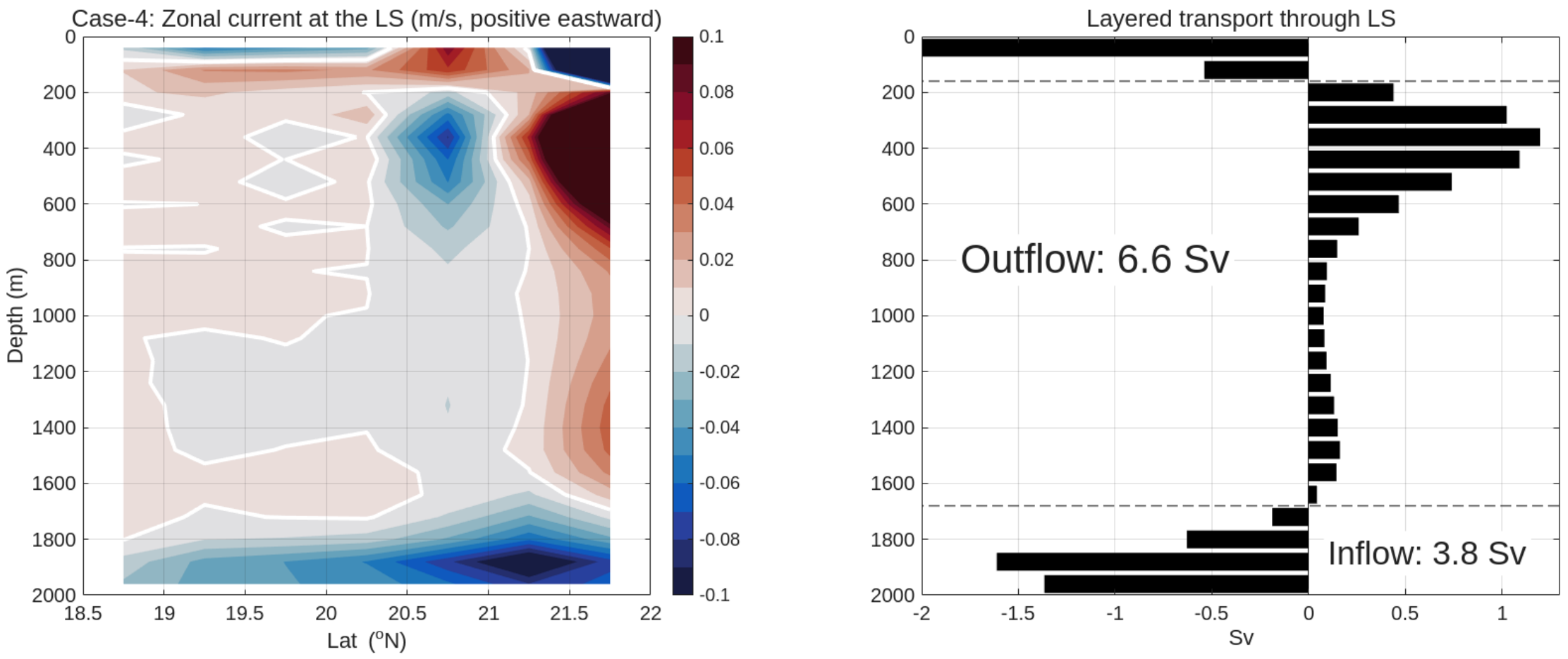

Fig. 10. Same as Fig. 3, but for Case-4.

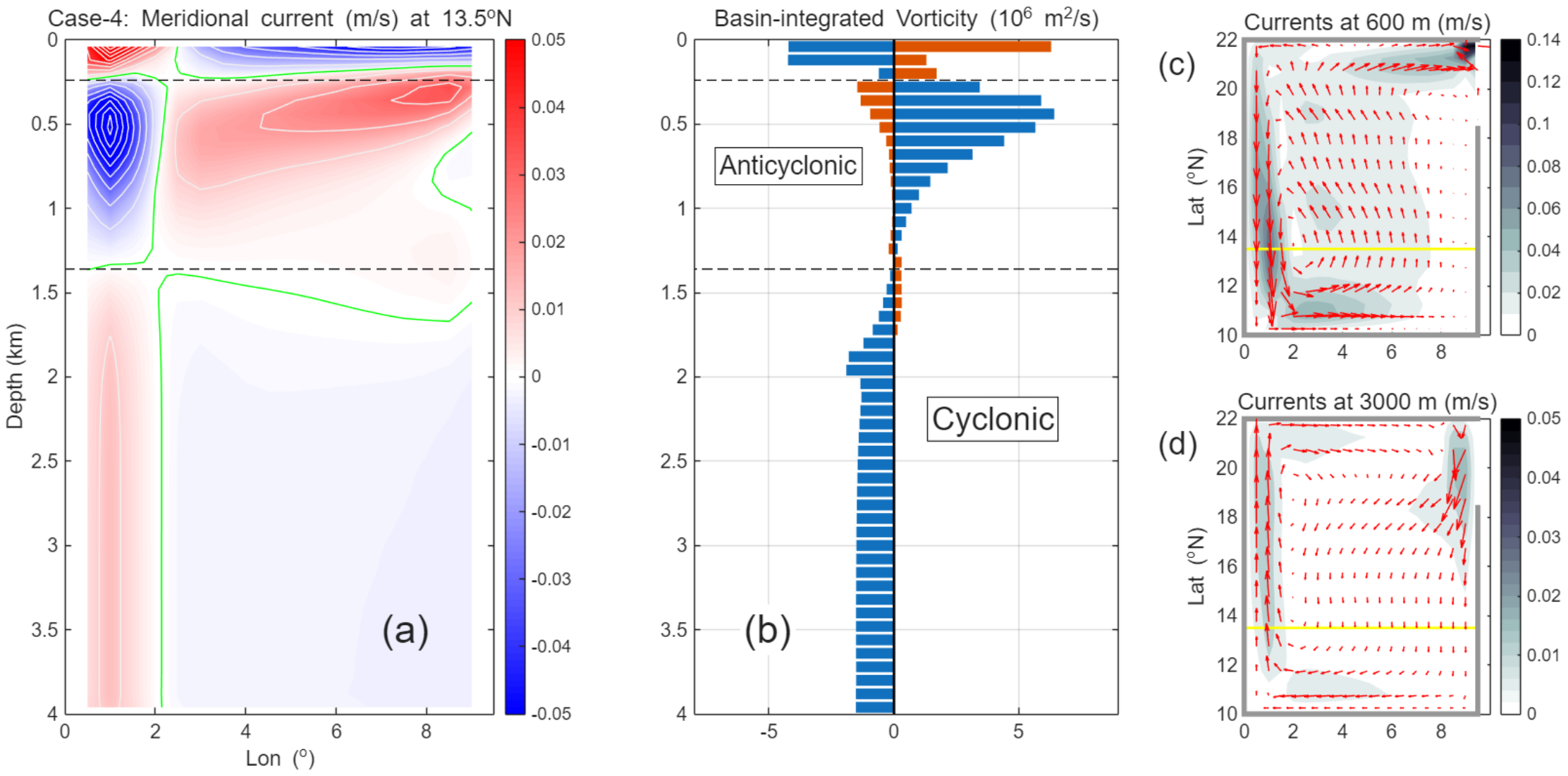

Fig. 11. Same as Fig. 4, but for Case-4.

Table 1 summarizes key circulation metrics for all experiments, including LS transport, WBC index, basin-integrated vorticity, and MOC index for the middle and deep layers. Layer-wise LS transports are derived by accumulating the vertical transport profiles shown in Figs. 3, 6, 8, and 10. WBC transport is calculated by integrating the poleward (or equatorward) transport within the western boundary region (150 km from the coast according to the model output) at each latitude. The WBC index is defined as the maximum WBC transport value across all latitudes; for Case-3, the southern boundary current is used.

Basin-integrated vorticity is computed by integrating tangential velocity along the SCS basin boundary, as described in Section 3.1.2. To reduce the influence of strong recirculation near the LS in the sensitivity experiments, the integration domain excludes the area within 100 km of the eastern boundary, as indicated by the green dashed box in Fig. 4c, d. This localized recirculation may be associated with the abrupt change in the Coriolis parameter across the LS, as it does not appear in the control run. For the control run, basin-integrated vorticity is comparable between the green box area and the full domain (not shown).

The MOC index is defined as the value of the Eulerian overturning streamfunction at 18°N and 2,000 m depth (Fig. 5), representing the net vertical volume flux across the 2,000 m level over the entire SCS south of 18°N.

**Table 1**. Summary of circulation metrics for the four numerical experiments in this study. Positive (negative) LS transport denotes outflow (inflow). Positive (negative) WBC index denotes northward (southward) transport for Cases 1, 2, and 4, and eastward (westward) transport for Case 3. Positive (negative) vorticity corresponds to counterclockwise (clockwise) circulation. The MOC index, defined at 2,000 m, is listed as deep-layer metric.

| Layer | Case ID | LS transport (Sv) | WBC index (Sv) | Basin-integrated vorticity ($10^6$ $m^2/s^2$) | MOC index (Sv) |
|---|---|---|---|---|---|
| Middle | 1 | 5.6 | 6.6 | –3.8 | – |
| | 2 | 6.3 | 6.6 | –3.9 | – |
| | 3 | 6.6 | –5.0* | –2.8 | – |
| | 4 | 6.6 | –4.5 | 2.3 | – |
| Deep | 1 | –3.0 | –6.8 | 2.8 | 1.8 |
| | 2 | –3.5 | –6.5 | 2.5 | 2.1 |
| | 3 | –4.1 | 3.4* | 0.82 | 2.4 |
| | 4 | –3.8 | 4.3 | –1.4 | 2.8 |

* Refers to SBC rather than WBC.

Beyond the shared features, each sensitivity case exhibits distinct characteristics shaped by its prescribed Coriolis configuration. In Case 3, where the $\beta$-effect is oriented zonally, the strongest boundary current system shifts from the western to the southern boundary—consistent with the imposed zonal gradient of planetary vorticity. In Case 4, although the sign of basin-integrated vorticity reverses relative to the other cases, the CAC circulation pattern persists owing to its negative Coriolis parameter.

The meridional hotspot of throughflow at the LS is modulated by the spatial distribution of planetary vorticity near the strait. A comparison among Cases 2–4 (Figs. 6, 8, 10) reveals that both the deep inflow and middle-layer outflow tend to occur on the side with weaker $f$. Additionally, comparison between Cases 1 and 2 shows that while the middle-layer outflow remains concentrated on the southern tip, the deep inflow hotspot shifts from the northern to the southern tip of the strait—an outcome attributable to the removal of planetary vorticity in the Pacific domain in Case 2.

In addition, the deep-layer vorticity dynamics in our sensitivity experiments challenge the vorticity-driven hypothesis of Yang and Price (2000), as outlined in Section 1. Their study asserts that "a downwelling source represents zero net production of PV. Thus, in a steady circulation driven solely by such a source, the basin-wide production of PV by frictional processes must also be zero" and, for a closed-basin model, "the vorticity production by friction must vanish" (see Eq. [15] in Yang and Price 2000). However, our simulations demonstrate that in the deep layer below 2,000 m, effectively configured as a closed basin without lateral open boundaries, the downwelling sourced from the LS deep inflow leads to nonzero basin-wide vorticity production through lateral friction at all depths. This results from the nonzero line integral of tangential velocity around the basin (see Table 1).

To summarize, our sensitivity experiments do not support the hypothesis that planetary vorticity influx across the LS is the primary driver of the CAC circulations in the SCS. One might argue that the planetary vorticity outflux is nonzero in the middle layer; while this is true, the deep-layer influx is explicitly removed by design. Nevertheless, the basin-integrated vorticity in the deep layer remains comparable in magnitude to that in the control run for Case 2, and substantial in Cases 3 and 4 (Table 1). These results suggest that the deep-layer cyclonic circulation can be sustained without planetary vorticity input from the Pacific and that the vorticity within the SCS is unlikely to be sourced from the LS lateral boundary.

# 4. Analytic volume-flux-driven model for the SCS

The idealized experiments in Section 3 reproduce the layered transports through the LS and the CAC circulations within the SCS, even in the absence of planetary vorticity influx through the LS. The basin-integrated vorticities in the zero-vorticity-influx cases are comparable in magnitude to those in the control run, which includes lateral vorticity input from the Pacific. These results challenge the vorticity-flux-driven hypothesis outlined in Section 1. In this section, we examine whether the volume-flux-driven mechanism—based on the Stommel-Arons framework—can account for the observed model behaviors.

## 4.1 Stommel-Arons model in Cartesian coordinates

Given the limited spatial extent of the SCS, the Cartesian coordinate system is commonly adopted in models to analyze its dynamics (e.g., Wang et al 2006; Wang et al 2012; Wang et al 2018; Quan and Xue 2019; CCG23; this study). While the original Stommel-Arons model was developed for global-scale dynamics using spherical coordinates, we reformulate it for a rectangular basin in Cartesian coordinates to facilitate its application to regional systems such as the SCS.

Consider a steady, vertically uniform circulation in a homogeneous layer of constant depth $H$, confined within a rectangular basin of zonal and meridional extents $L_x$ and $L_y$, respectively. The linearized momentum equations under Cartesian coordinates are:

$$-fv = -\frac{1}{\rho}\frac{\partial p}{\partial x}\,, \tag{4}$$

$$fu = -\frac{1}{\rho}\frac{\partial p}{\partial y}\,, \tag{5}$$

where $u$ and $v$ are the velocity component in the zonal ($x$) and meridional ($y$) directions, $p$ is pressure, and $\rho$ is the reference water density. Under the $\beta$-plane assumption, the Coriolis parameter is given by:

$$f = f_0 + \beta y\,, \tag{6}$$

where $f_0$ is the Coriolis parameter at the southern boundary, and $\beta$ is a constant parameter.

Assume a point source or sink $S_0$ introduces a volume flux $Q$ to the layer ($Q > 0$ for a source; $Q < 0$ for a sink). Following Stommel and Arons (1959), the vertical velocity $w_e$ induced by $Q$ is assumed to be uniformly distributed, yielding the continuity equation,

$$\frac{\partial u}{\partial x} + \frac{\partial v}{\partial y} = -\frac{w_e}{H}. \tag{7}$$

where $w_e$ is given by,

$$w_e = \frac{Q}{L_x L_y}. \tag{8}$$

In the context of the SCS middle layer, $w_e$ represents the net vertical exchange across both layer interfaces.

Eliminating pressure from Eqs. (4) and (5) yields the PV balance,

$$\beta v = f\frac{w_e}{H}. \tag{9}$$

which expresses the conservation of PV in the presence of vertical motion, with the $\beta v$ term balancing vortex stretching or compression (Yuan 2002; Wang et al 2018).

Applying the no-penetration boundary condition $u(x = L_x) = 0$ to Eq. (4) implies that $p(x = L_x, y)$ is constant, which we set to zero without loss of generality. Integrating Eq. (4) westward then yields the pressure field,

$$\frac{p}{\rho} = \frac{(f_0 + \beta y)^2}{\beta H} w_e(x - L_x)\,. \tag{10}$$

Substituting Eq. (10) back into Eqs. (4)–(5) provides analytical solutions for the velocity components,

$$u = -\frac{w_e}{H}(x - L_x), \qquad v = \frac{w_e}{H}\left(y + \frac{f_0}{\beta}\right), \tag{11}$$

Using parameters consistent with Case-1:

- $L_x = 900\ km,\ L_y = 1200\ km,$
- $f_0 = 2.25 \times 10^{-5} s^{-1},\ \beta = 2.2 \times 10^{-11} m^{-1} s^{-1},$
- $Q_{mid} = -5.6$ Sv, $Q_{bot} = 3.0$ Sv, (from Fig. 3 or Table 1)
- $H_{mid} = 1080\ m,\ H_{bot} = 2680\ m.$ (from Fig. 4)

Fig. 12 presents the resulting pressure fields ($p/\rho$) in the middle and deep layers,

respectively.

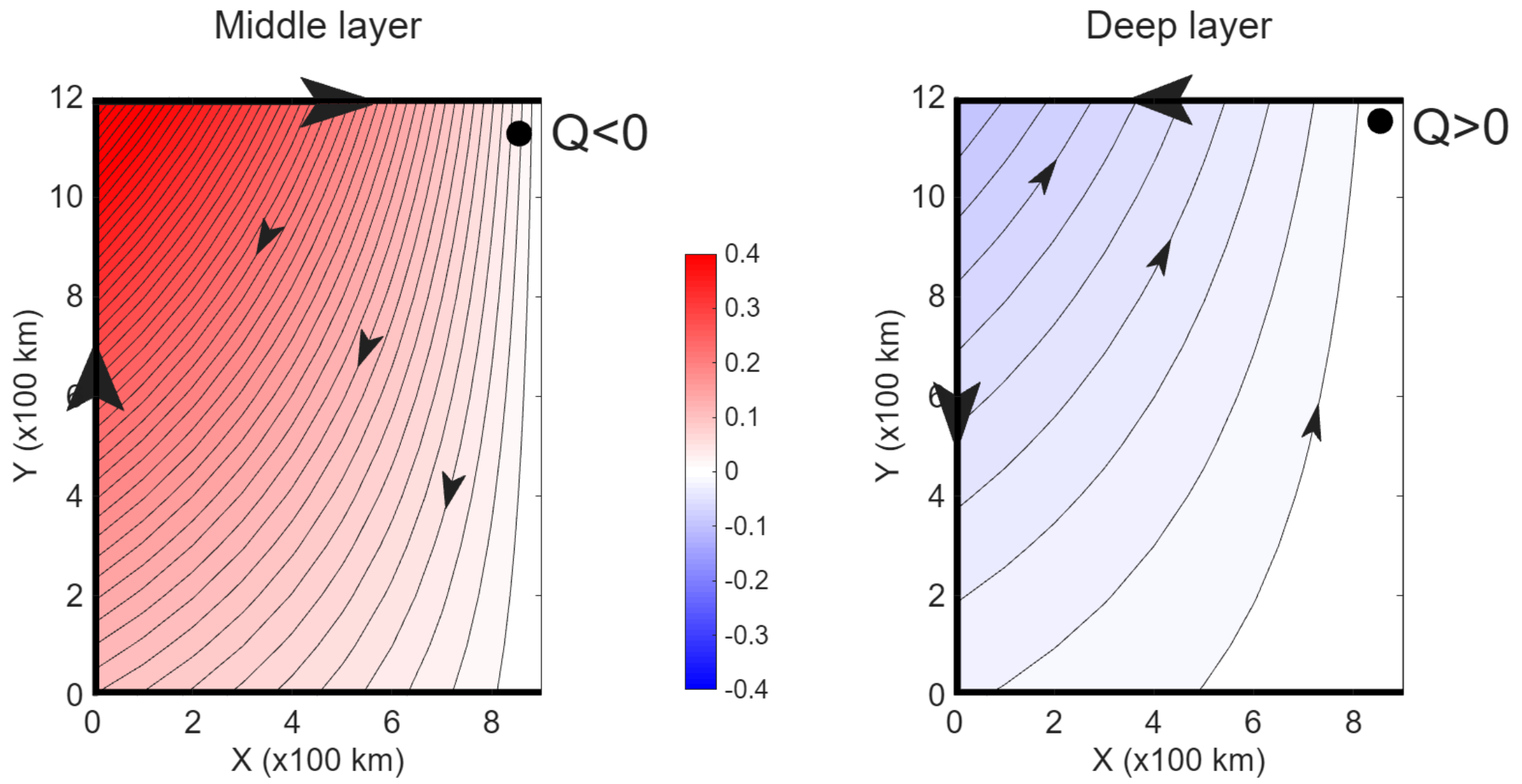

Fig. 12. Stommel-Arons model solutions for pressure ($p/\rho$, contours, unit: $m^2/s^2$) in the middle layer (left) and deep layer (right) of the SCS basin, based on Case-1 parameters. Small arrows indicate interior flow directions, while large arrows denote boundary currents directions. The positions of the source/sink influence the calculation of boundary current transports (see Section 4.2). (CI: 0.01 $m^2/s^2$ in both panels)

Though inherently simple, the Stommel-Arons model, when applied to rectangular basin, captures several key features consistent with Case-1, including:

1) **Basin-integrated vorticity**: The analytic model generates an anticyclonic circulation in the middle layer and a cyclonic one in the deep layer, effectively replicating the alternating-layer structure.
2) **Stronger interior flow in the middle layer**: The pressure gradients (Fig. 12) imply more vigorous interior flow in the middle layer compared to the deep layer, consistent with the model output in Case-1 (Fig. 4).
3) **Northern and southern boundary currents (NBC and SBC)**: In addition to the WBC, Case-1 reveals the presence of northern and southern boundary currents (NBC and SBC). Although the deep NBC has been reported in both observational and modeling studies (e.g., Lan et al 2013; Shu et al 2014; Zhou et al 2020), its theoretical foundation has remained elusive. The analytical solution presented here (Fig. 12) successfully reproduces this feature, demonstrating strong agreement with the model results (Fig. 4c, d).

The directions of these boundary currents can be inferred from volume conservation following Stommel and Arons (1959), either qualitatively or quantitatively, as elaborated in subsequent sections.

Given its simplicity, the Stommel–Arons framework proves remarkably effective in capturing essential qualitative aspects of the SCS circulation. In the following section, we extend this analysis through a quantitative comparison with model output.

**4.2 Comparison of WBC transports**

This section compares the WBC transports predicted by the Stommel–Arons analytical solution with the model results from Case-1. Within the theoretical framework, WBC transport is estimated using volume conservation principles. For the deep-layer WBC at a given location $y$, denoted as $T_w(y)$, we consider a subdomain bounded by $x=0$, $x=L_x$, and a zonal section at $y$ (Fig. 13). The total outward volume flux consists of two components: the interior transport ($T_{int}$) and vertical (upwelling) transport due to the point source ($T_{up}$) (Fig. 13). The WBC ($T_w$) represents the sole inward volume flux into this subdomain. Applying volume conservation yields,

$$T_w(y) + \int_0^{L_x} vHdx + w_e y L_x = 0\,, \tag{12}$$

Substituting Eqs. (8) and (11) gives (positive $Q$ for source),

$$T_w(y) = -\frac{2y + f_0/\beta}{L_y} Q\,. \tag{13}$$

This expression provides the theoretical WBC transport for a source/sink-driven circulation in a rectangular basin. It applies to the middle layer as well, with $Q$ being negative, resulting in a poleward (positive) WBC transport (Fig. 12). Note that this result is invalid if the source or sink lies within the bounded subdomain.

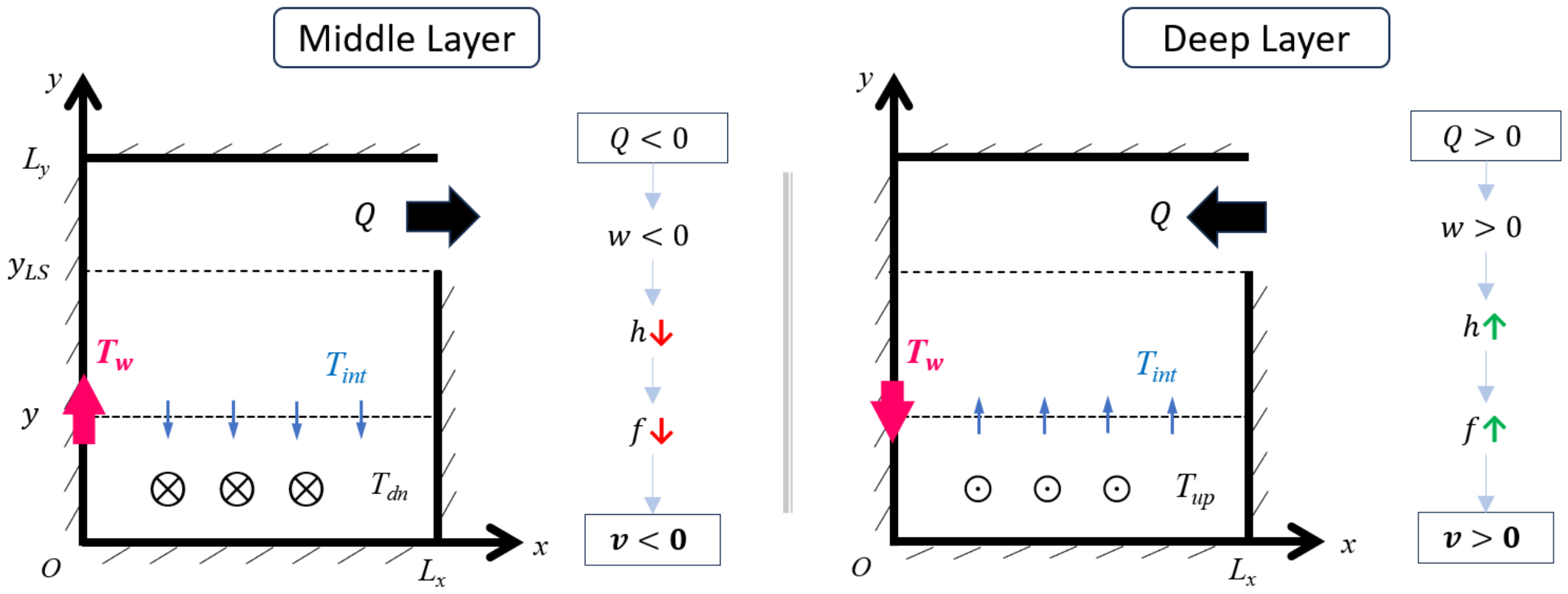

Fig. 13. Schematic diagram illustrating the calculation of the WBC transport in the middle layer (with a sink, left) and the deep layer (with a source, right) in Case-1. The dynamic relationship whereby a source (sink) drives poleward (equatorward) interior flow in the deep (middle) layer of the basin governed by conservation of PV (*f/h*) is also presented.

To apply the theoretical solution to the model basin discussed in the previous section, two modifications are necessary:

1) The wide-open LS cannot be treated as a point source or sink, as assumed in the theoretical framework. Therefore, Eq. (13) is applied not to the full SCS domain but to a subdomain south of the LS, bounded meridionally at 18°N (i.e., equatorward of 18°N). The subdomain has a meridional extent of 850 km (denoted as $y_{LS}$ in Fig. 13), corresponding to the region from $y = 0$ to $y = y_{LS}$ in the schematic (Fig. 13).
   The volume fluxes out of the middle layer and into the deep layer within this subdomain are derived from the Eulerian MOC streamfunction in Fig. 5. At 18°N, the streamfunction values at the interface depths (approximately 240 m and 1,320 m; see Fig. 4) are about −1.5 Sv and 2.0 Sv, respectively, indicating an outflow of 3.5 Sv from the middle layer ($Q_{middle} = -3.5$ Sv) and an inflow of 2.0 Sv into the deep layer ($Q_{deep} = 2.0$ Sv).
2) Because of the limited zonal extent of the SCS basin, the WBC occupies a significant width relative to the interior flow. Thus, the full zonal width should not be assigned to the interior region. Based on Fig. 4, the WBC spans approximately 150 km, while the interior flow occupies the remaining 750 km. Accordingly, a reduced width of 750 km (denoted as $L_x^{int}$) is used for the integration in Eq. (12), instead of the full basin width of 900 km ($L_x$).

With these adjustments, the modified theoretical WBC transport is expressed as:

$$T_w(y) = -\frac{Q}{L_x y_{LS}}\left[\frac{f_0}{\beta}L_x^{int} + y\left(L_x + L_x^{int}\right)\right]. \tag{14}$$

For comparison with Case-1 model output, the middle-layer WBC transport is computed by integrating all poleward transport within 150 km of the western boundary at each zonal section. For the deep layer, all equatorward transport below 500 m within the same offshore range is integrated.

Figure 14 presents the comparison between the theoretical predictions [Eq. (14)] and the modeled WBC transports for both the middle and deep layer. The simplified theoretical solutions produce transport magnitudes comparable to the model results and successfully captures their strengthening trend with latitude. The agreement is particularly striking in the deep layer, where the theoretical prediction closely matches the latitudinal linear variation of the modeled transports.

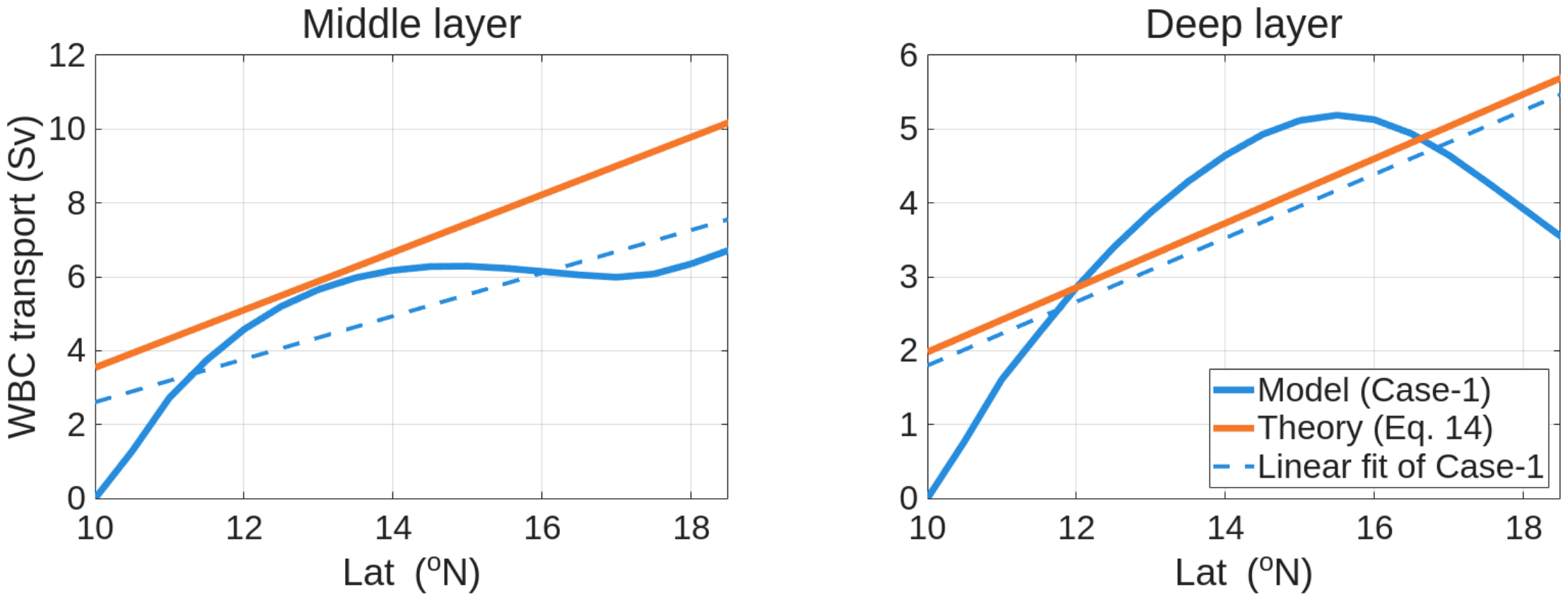

Fig. 14. Comparison of western boundary current (WBC) transports at different latitudes, showing the Stommel–Arons theoretical predictions (orange line, Eq. 14) versus the idealized model output (solid blue curve). Results are shown for the middle layer (left) and deep layer (right). Dashed lines indicate linear fits to the modeled WBC transports as a function of latitude.

### 4.3 Stommel-Arons solutions for Case-4

To further assess the applicability of the Stommel–Arons framework in explaining the dynamics of the layer-dependent circulation, we apply the analytical solution using the Coriolis parameter configuration (Fig. 2) and LS transports (from Fig. 6 or Table 1) from the idealized experiment, Case-4, in which the Coriolis parameter is negative. The results are shown in Fig. 15. The poleward flow in the middle-layer interior and the equatorward flow in the deep-layer interior align with the modeled circulations (Fig. 11).

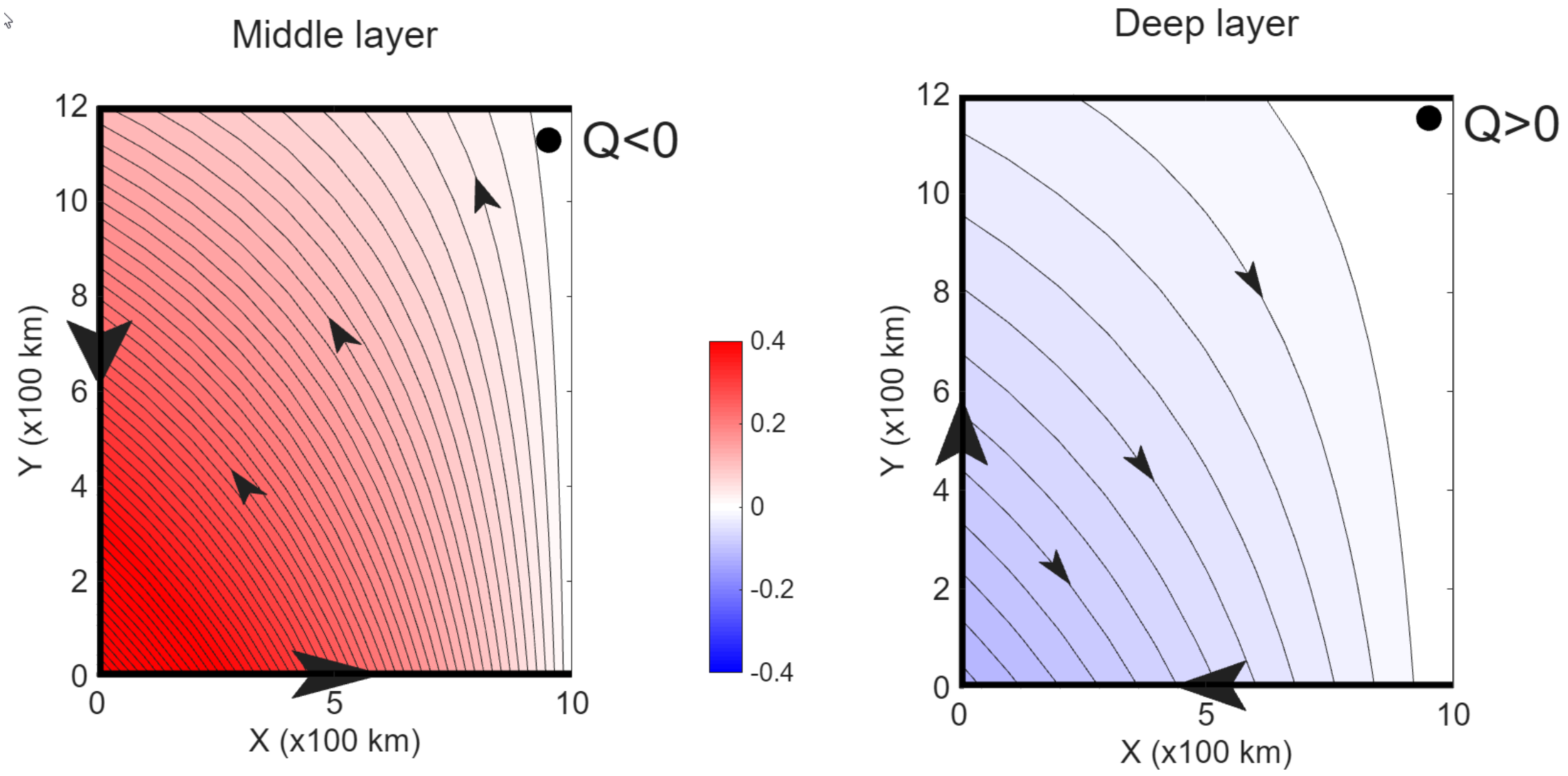

Fig. 15. Same as Fig. 12, but using the Coriolis parameter for Case-4. This Stommel–Arons solution is intended for comparison with the Case-4 model output shown in Fig. 11.

In contrast to the Stommel-Arons solution for Case-1 (Fig. 12), the solution for Case-4 (Fig. 15) exhibits a stronger pressure gradient along the southern boundary than along the northern boundary, suggesting a more pronounced SBC relative to the NBC. This pattern agrees with the model output, which also shows an intensification of the SBC relative to its northern counterpart in Case-4 (Fig. 11) relative to other cases (Fig. 4 or Fig. 7).

The direction of the SBC can be qualitatively inferred using the volume conservation framework (similar to the illustration in Fig. 13). Take the middle layer as an example: for a closed subdomain bounded by a given meridional section to the east, both the interior flow and the vertical transport are directed inward. Consequently, the combined transport of the SBC and NBC must be outward, i.e., eastward. Given the stronger pressure gradient along the southern boundary, the transport there dominates, resulting in an eastward SBC, consistent with the model output (Fig. 11).

However, a quantitative estimate is needed to determine the directions of the WBCs, since the interior flow becomes outward for the closed subdomains bounded by $y = 0$ and a zonal section $y$ within the range $0 < y < y_{LS}$, where $y_{LS}$ denotes the meridional coordinate of the southern end of the LS (Fig. 13). In Eq. (13), the parameter $f_0/\beta$ is approximately −2,200 km for Case-4, making the numerator term $2y + f_0/\beta$ negative for $y < 1{,}100$ km, and positive for $y > 1{,}100$ km. For a sink in the middle layer ($Q$<0, Fig. 15), this implies that the WBC should be directed southward for $y < 1{,}100$ km and northward for $y > 1{,}100$ km. Within the current basin geometry, this suggests a southward WBC along most of the boundary, with a northward reversal near the northwestern corner.

However, such a reversal is not observed in the model output (Fig. 11). This discrepancy may arise from limitations of the Stommel–Arons solution or, more likely, from the breakdown of the closed-subdomain assumption. In this region, the LS transport contributes to the subdomain's volume budget, thereby invalidating the assumptions underpinning Eqs. (12) and (13), as noted in section 4.2.

More generally, the analytical model has limitations. For example, for the configuration of Case-4, it predicts a weakening trend in WBC transport with increasing latitude—an outcome that contrasts with the results from the MITgcm simulation.

## 5. Summary and discussion

This study investigates two competing hypotheses concerning the formation mechanisms of the layered cyclonic–anticyclonic–cyclonic (CAC) circulation system in the South China Sea (SCS) through idealized numerical experiments using the MITgcm. We developed a model that replicates the CAC circulation pattern in the SCS with one of the simplest configurations documented in the literature. This minimal yet effective setup isolates the key dynamical processes governing the layer-dependent structure. The model is thermodynamically forced by diffusivity contrast between the two basins and achieves a steady state through temperature relaxation at the outer boundaries of the Pacific domain. Its reliability is validated by comparisons with a higher-resolution counterpart and an expanded full-width Pacific basin domain.

Sensitivity experiments, which impose zero planetary vorticity in the Luzon Strait (LS) and the Pacific, demonstrate that the layered structure and the intensity of the SCS circulations remain largely unaffected. These findings suggest that the CAC circulation is not primarily driven by planetary vorticity flux through the LS, challenging the vorticity-driven hypothesis.

By adapting the Stommel-Arons theoretical framework from spherical to Cartesian coordinates, we derived analytical solutions for the SCS circulation, incorporating a point volume source/sink in the northeast corner. Despite its simplicity, this analytical model accurately captures key features consistent with the MITgcm output, both qualitatively and quantitatively. These include the direction of the interior flow, the intensity contrast between the middle and deep layers, and the direction and latitudinal variation of boundary current transports. Together, these results offer robust support for the volume-driven hypothesis.

Overall, this study challenges the hypothesis that the CAC circulation in the SCS is primarily driven by planetary vorticity influx through the LS or by flow–topography interactions. Instead, it supports a volume-driven mechanism associated with throughflows across the LS. However, we argue that the CAC circulation system is not generated by throughflows in both layers. Rather, only the deep inflow serves as the driving force, while the middle-layer outflow is not a cause but a consequence of volume conservation.

The underlying mechanism is as follows (Fig. 13): deep inflow through the LS enters the deep layer of the SCS, inducing vertical stretching of the deep-layer water column. By continuity, this stretching compresses the overlying middle layer. Through PV conservation, these vertical deformations on water columns generate opposing meridional interior flows in the deep and middle layers. Together with returning WBCs, these flows establish cyclonic and anticyclonic circulations in the respective layers. Thus, the deep inflow through the LS acts as the sole driver in this process, with the middle-layer outflow emerging as a dynamical consequence of middle-layer compression.

While the idealized model developed in this study suggests that volume-flux forcing may play a key role in shaping the layer-dependent circulations in the SCS, we do not exclude the influence of other factors or mechanisms. The Stommel-Arons solution also has some limitations in certain aspects. The potential contributions of additional processes remain open for future investigation. Beyond exploring the fundamental dynamics, this study provides a useful baseline for future examinations of how external factors—such as diffusivity, topography, surface forcing, and the presence of the Kuroshio—affect the water exchange across the LS and layer-dependent pattern of the SCS circulations.

**Acknowledgments**

This work was supported by NSFC grant 42476006 and grant XMUMRF/2024-C14/ICAM/0017 from the Xiamen University Malaysia Research Fund.

# Drivers of Layered Circulations in the South China Sea: Volume Flux or Vorticity Flux?

Lei Han [a, b]

*[a] China-ASEAN College of Marine Sciences, Xiamen University Malaysia, Sepang, Malaysia*

*[b] College of Ocean and Earth Sciences, Xiamen University, Xiamen, China*

*Corresponding author*: Lei Han, lei.han@xmu.edu.my

## Supplementary materials

This document includes

- Two supplementary tables, from Table S1 to Table S2;
- Five supplementary figures, from Fig. S1 to Fig. S6;

Table S1. Model parameters used in the MITgcm simulations conducted in this study.

| Parameters<br>(MITgcm variable) | Value / Setting |
|---|---|
| Coordinate system | Cartesian |
| Model domain | SCS domain: 10°N-22°N, 0-10°E<br>Pacific domain: 0°-35°N, 10-40°E<br>(Full-width Pacific also tested [Table S2]) |
| Horizontal resolution (delX, delY) | 50 km<br>(10-km resolution also tested [Table S2]) |
| Vertical resolution (delR) | 80 m |
| Basin depth | 4000 m (SCS and Pacific)<br>2000m (Luzon Strait) |
| Temperature restoring and timescale | See Fig. 1 |
| Wind stress | None |
| Time step (deltaT) | 30 min<br>(60-min time step tested in Table S2) |
| EOS type (eosType) | Linear |
| Thermal expansion coefficient (tAlpha) | $2\times10^{-4}$ °C$^{-1}$ |
| haline contraction coefficient (sBeta) | 0 |
| GM package | enabled |
| Horizontal viscosity (viscAh) | $2\times10^{5}$ m$^2$/s |
| Diapycnal diffusivity (diffKrFile) | $1\times10^{-3}$ m$^2$/s for the SCS and LS<br>$1\times10^{-5}$ m$^2$/s for the Pacific |
| Vertical viscosity (viscAr) | $1.2\times10^{-4}$ m$^2$/s |
| Bottom drag coefficient (bottomDragLinear) | $5\times10^{-4}$ m/s |
| Coriolis parameter | Four configurations as shown in Fig. 2 |
| Beta-plane parameter (beta) | $2.2\times10^{-11}$ m$^{-1}$s$^{-1}$ |

Table S2. Same as Table 1, but for sensitivity tests assessing the robustness of the control run (Case-1), detailed as follows:

1) **Case-1/A (Testing equilibrium state)**: Mean values over model years 91–100, instead of 191–200 as in Case-1.
2) **Case-1/B (Testing time step)**: Same as Case-1, but with a doubled time step of 60 minutes instead of 30 minutes.
3) **Case-1/C (Testing Pacific width)**: Same as Case-1b, but with a full-width Pacific domain (120°E–100°W; see Fig. 1).
4) **Case-1/D (Testing horizontal resolution)**: Same as Case-1a, but with a 60-minute time step and a horizontal resolution of 10 km instead of 50 km.

| Layer | Case ID | LS transport (Sv) | WBC index (Sv) | Basin-integrated vorticity ($10^6$ $m^2/s^2$) | MOC index (Sv) |
|---|---|---|---|---|---|
| Middle | **1 (control)** | **5.6** | **6.6** | **–3.8** | - |
| | 1/A | 5.6 | 6.6 | –3.8 | - |
| | 1/B | 5.6 | 6.6 | –3.8 | - |
| | 1/C | 5.6 | 6.8 | –4.0 | - |
| | 1/D | 6.0 | 7.8 | –8.8 | - |
| Deep | **1 (control)** | **–3.0** | **–6.8** | **2.8** | **1.8** |
| | 1/A | –3.0 | –6.7 | 2.7 | 1.7 |
| | 1/B | –3.0 | –6.8 | 2.8 | 1.8 |
| | 1/C | –3.0 | –5.3 | 2.8 | 1.8 |
| | 1/D | –3.1 | –6.0 | 2.8 | 2.1 |

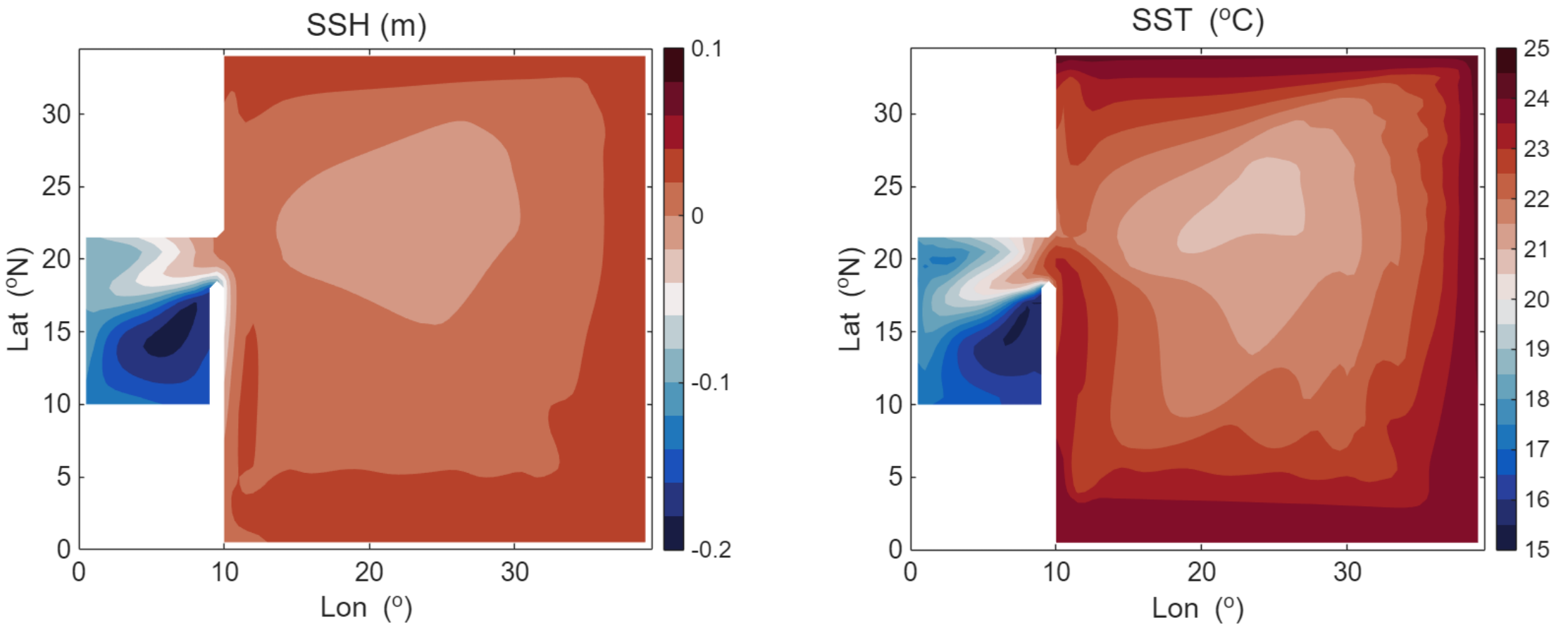

Fig. S1. Sea surface height (SSH; left) and sea surface temperature (SST; right) averaged over model years 191–200 for Case-1. The SSH and SST fields are fairly correlated, indicating the intrusion of Pacific water into the SCS.

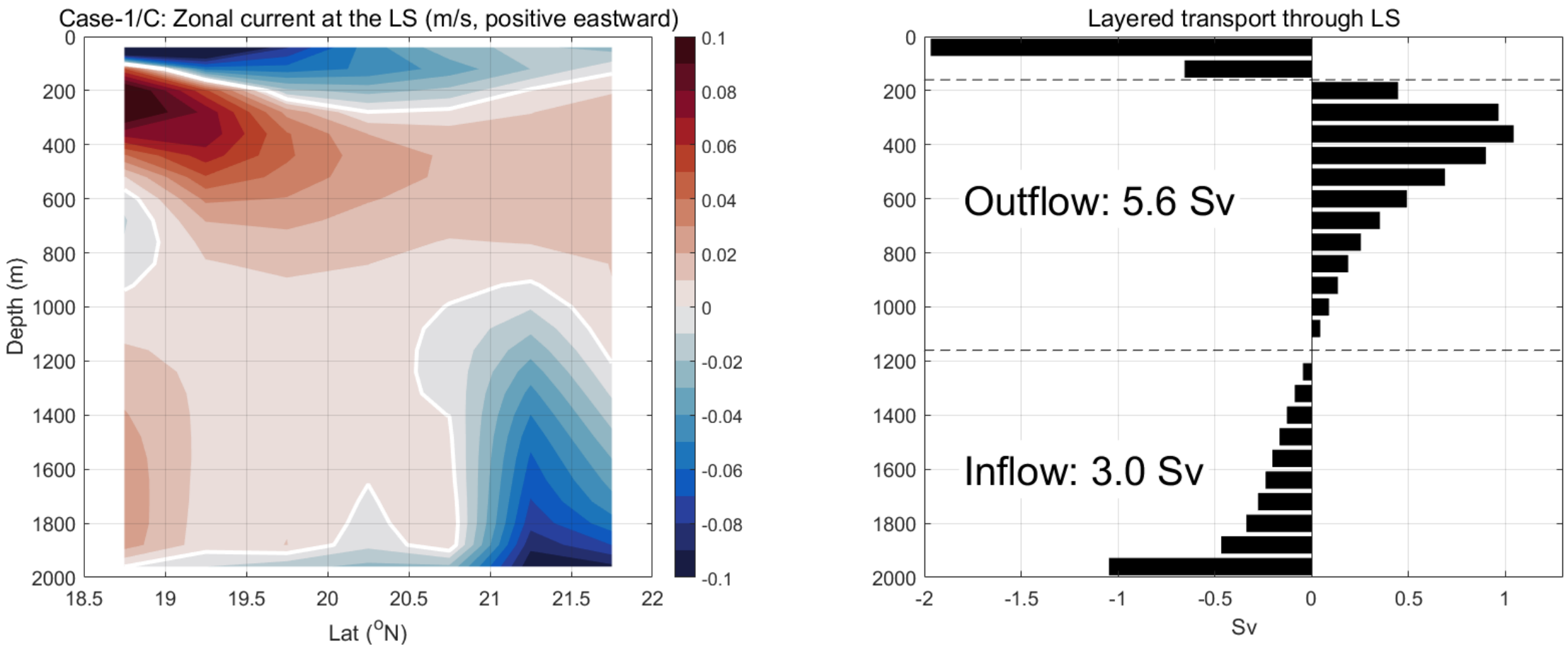

Fig. S2. Same as Fig. 3, but for Case-1/C, which uses full-width Pacific basin (Table S2).

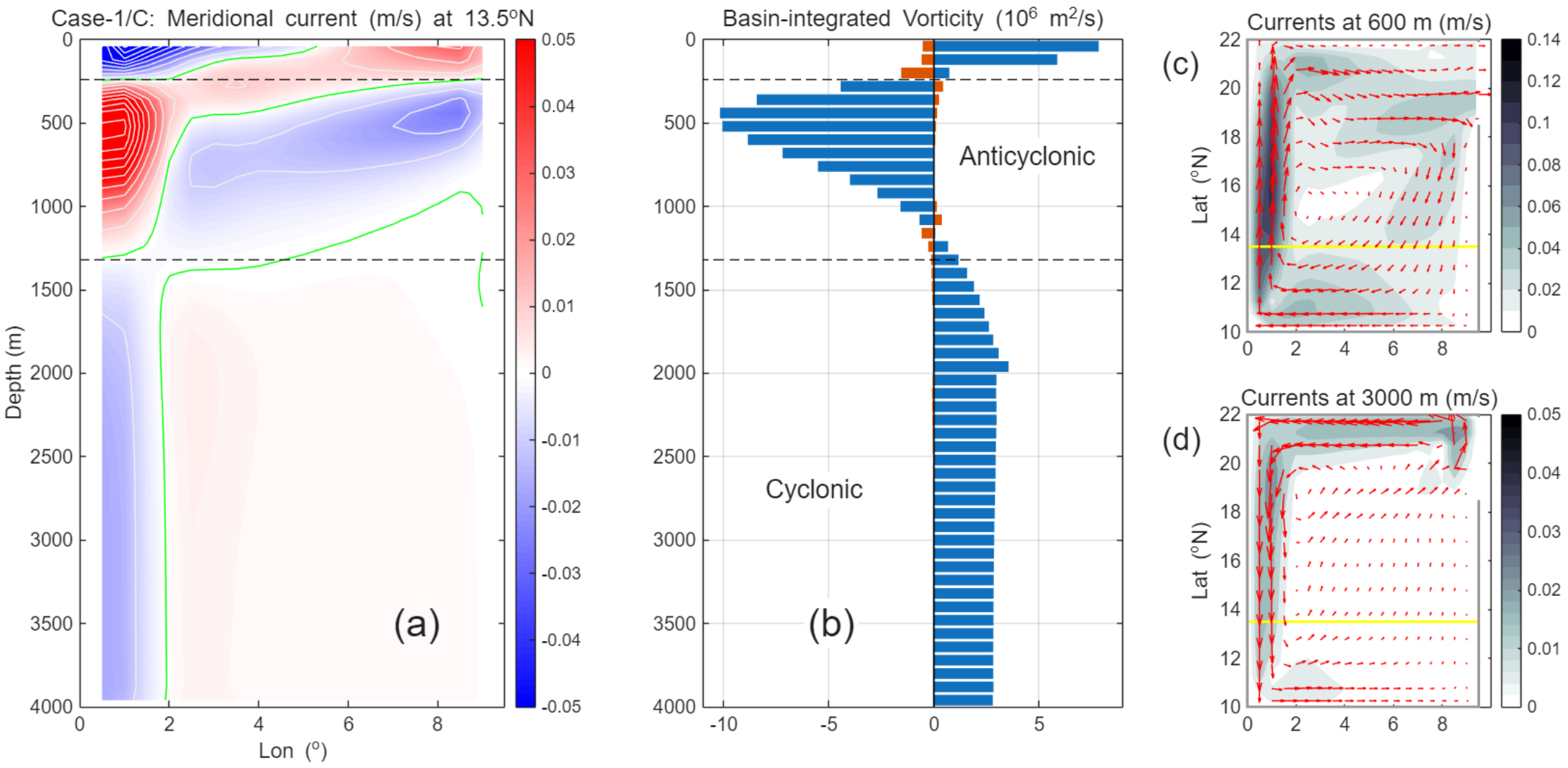

Fig. S3. Same as Fig. 4, but for Case-1/C, which uses full-width Pacific basin (Table S2).

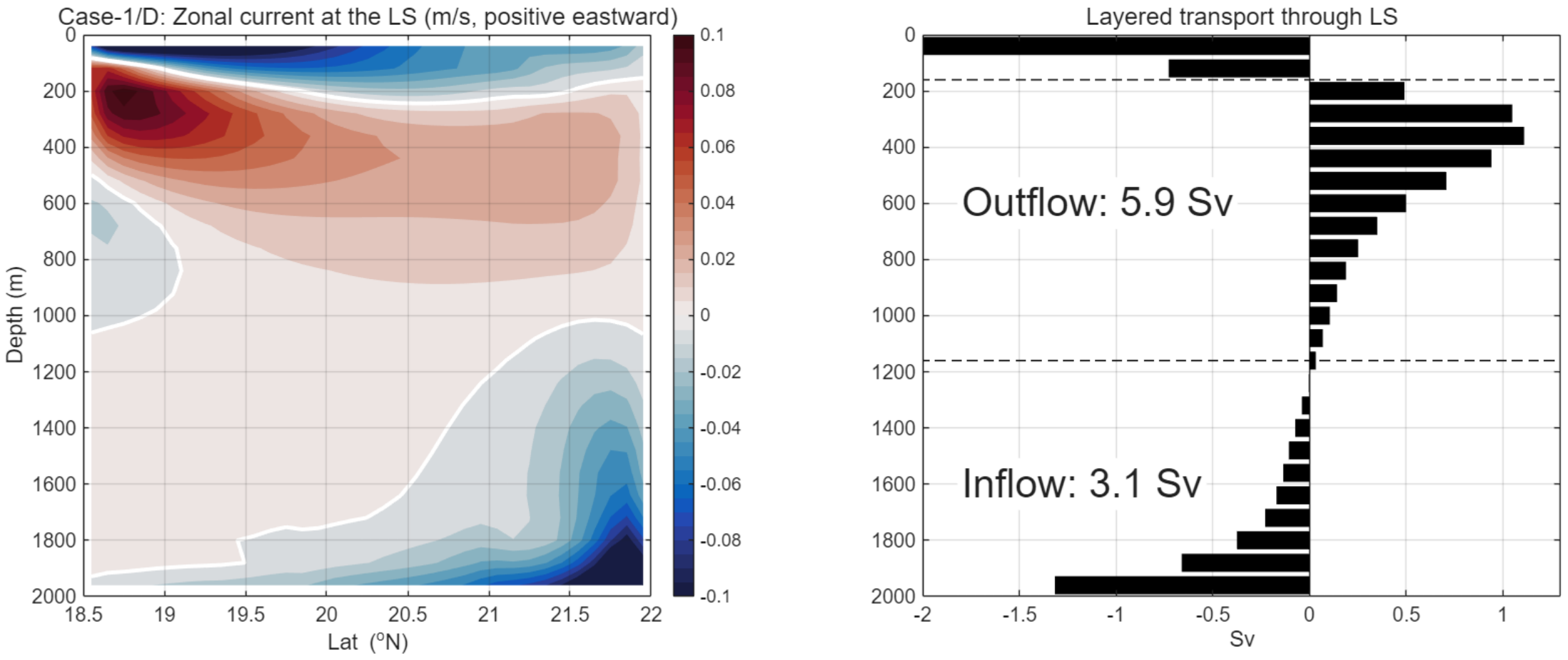

Fig. S4. Same as Fig. 3, but for Case-1/D, which uses a higher horizontal resolution of 10 km (Table S2).

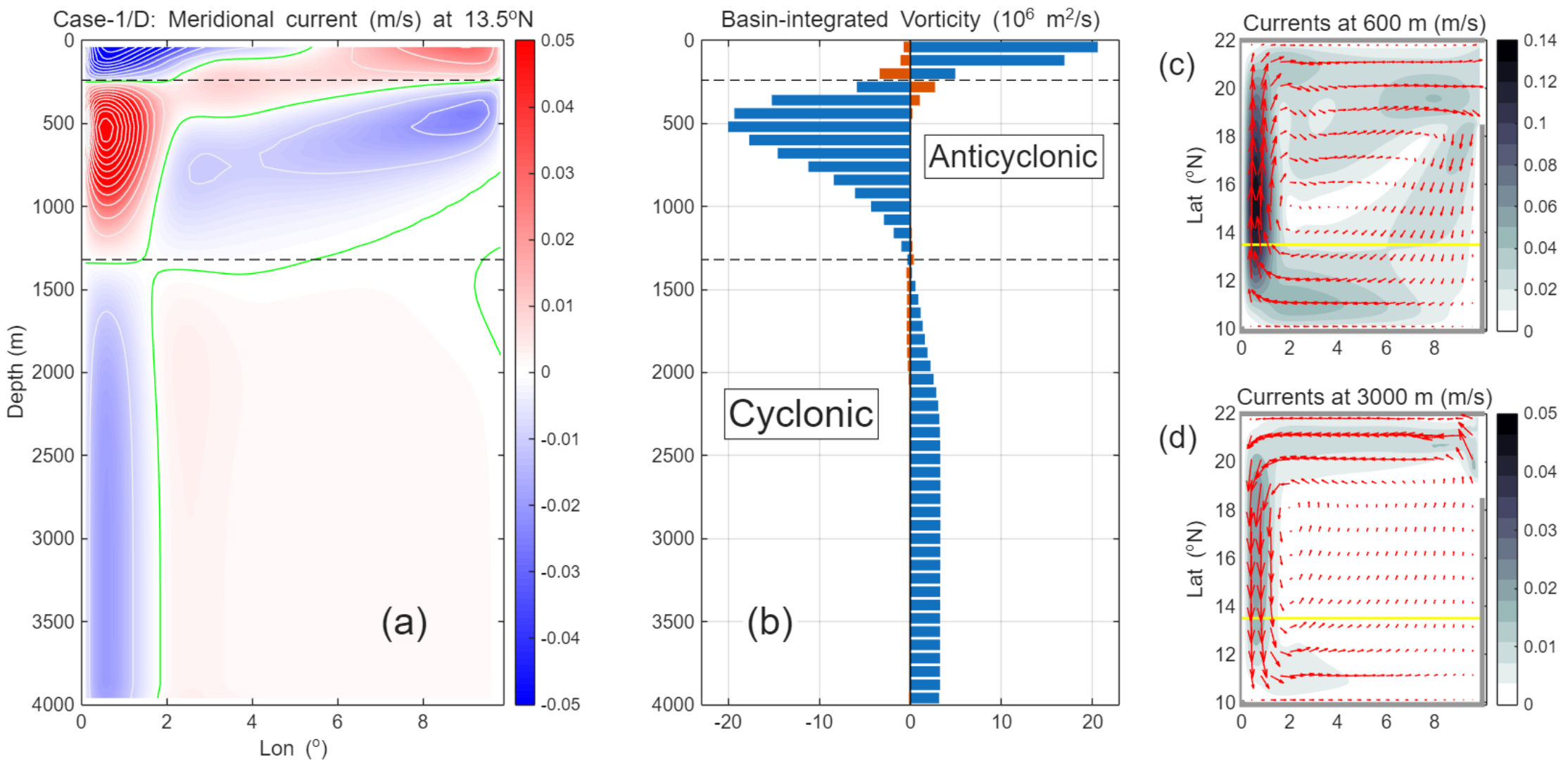

Fig. S5. Same as Fig. 4, but for Case-1/D, which uses a higher horizontal resolution of 10 km (Table S2).

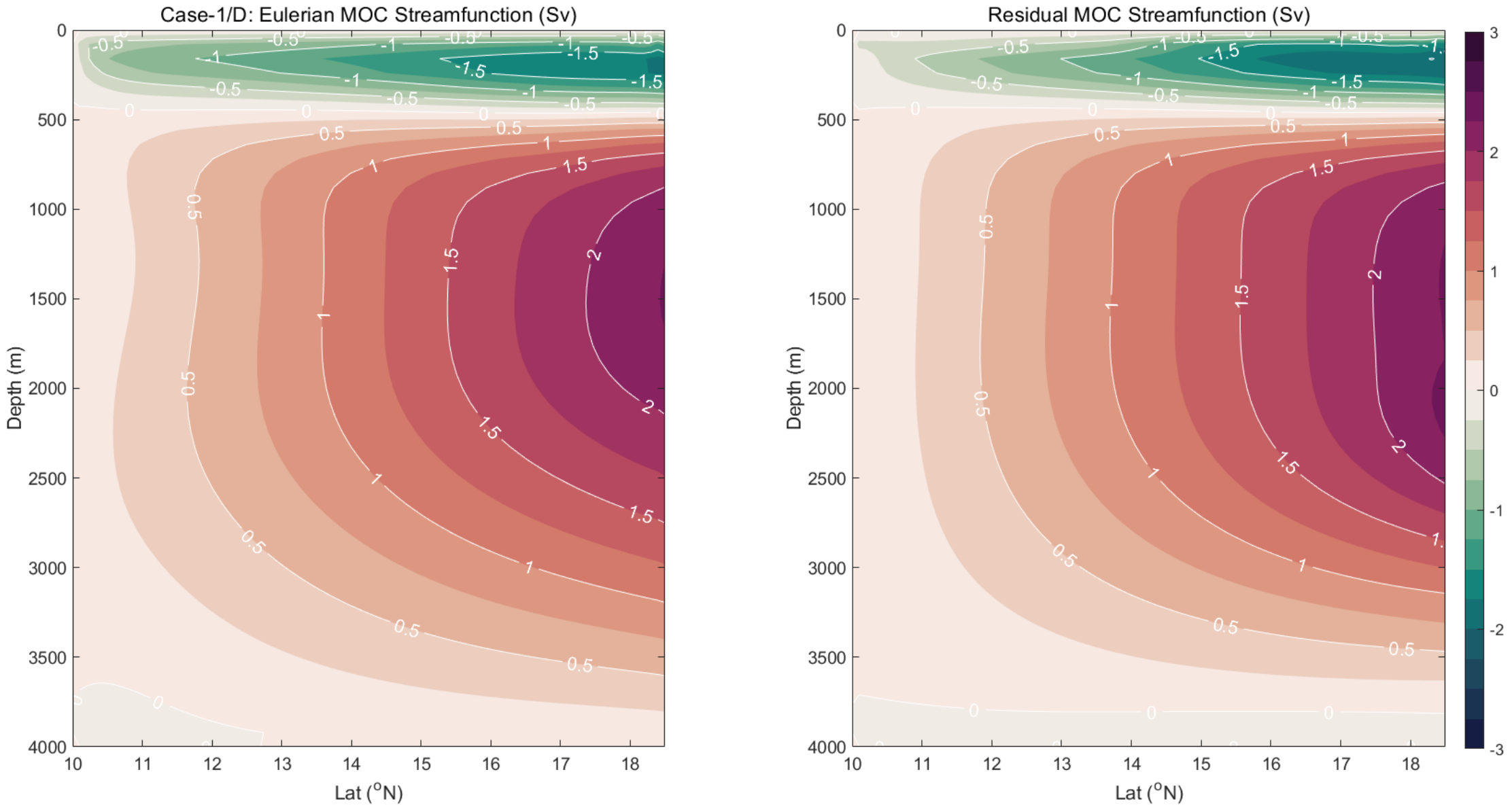

Fig. S6. Same as Fig. 5, but for Case-1/D, which uses a higher horizontal resolution of 10 km (Table S2).